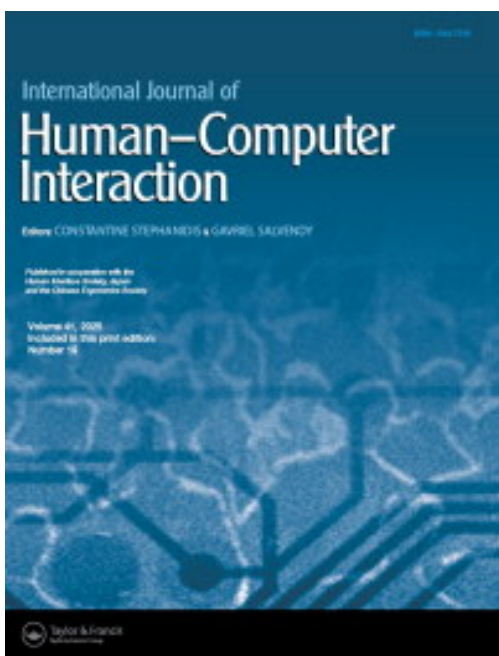

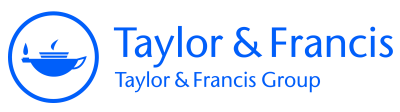

# COLP: Scaffolding Children's Online Long-Term Collaborative Learning

Siyu Zha, Yuanrong Tang, Jiangtao Gong & Yingqing Xu

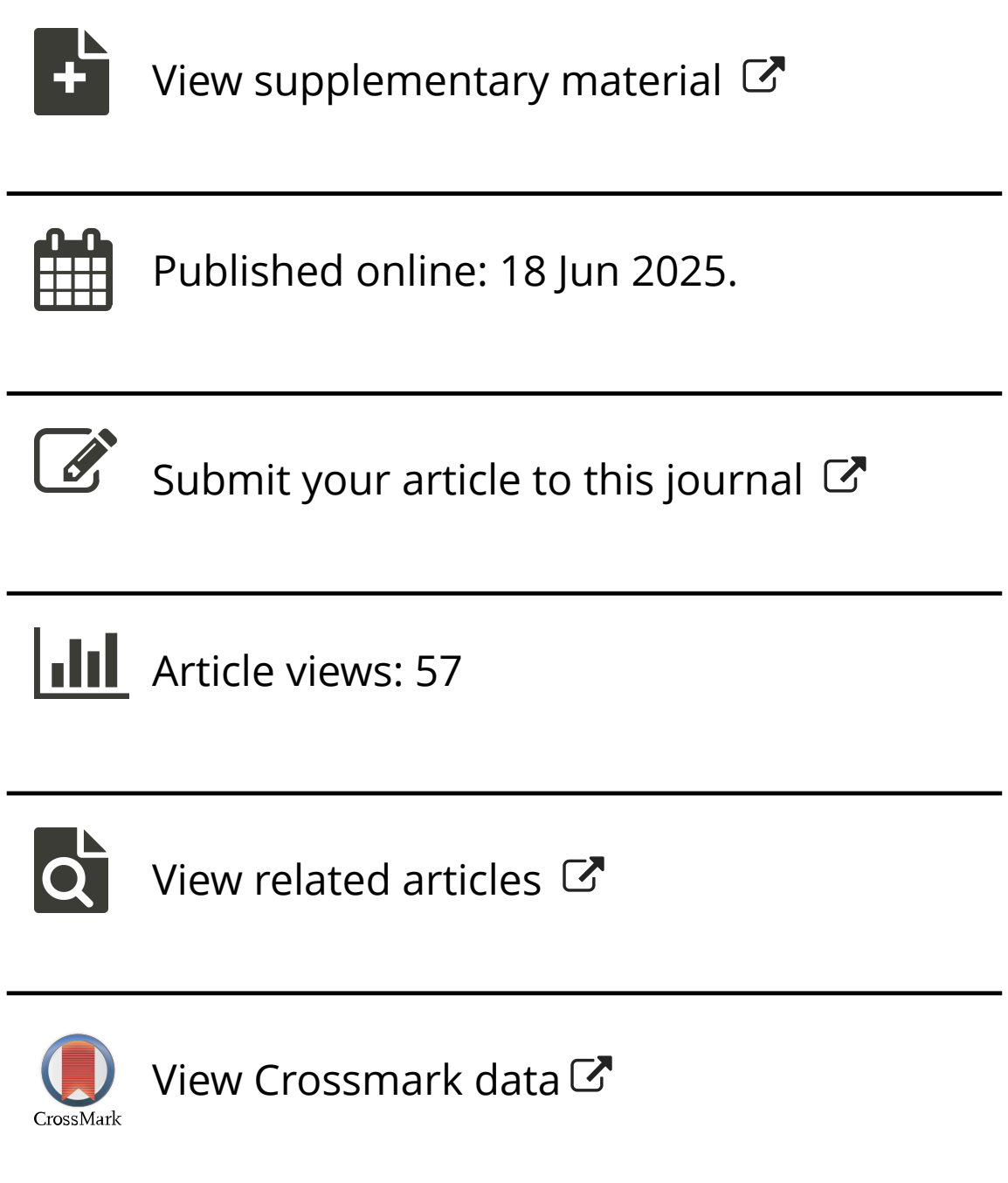

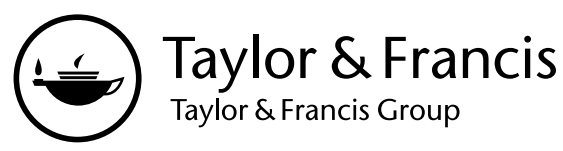

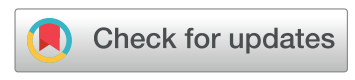

# COLP: Scaffolding Children's Online Long-Term Collaborative Learning

Siyu Zha, Yuanrong Tang, Jiangtao Gong, and Yingqing Xu

Tsinghua University, Beijing, China

**ABSTRACT**

Online collaborative learning is increasingly important, yet children still face challenges communicating and working together virtually, limiting their engagement in long-term teamwork. To address this, we designed the Children's Online Long-term Program (COLP), a 16-week online project-based learning program grounded in multiple learning theories. The program was implemented with 67 upper primary school students (Grades 3–6, ages 8–13) across five provinces in China. Results show that over one-third of participants sustained engagement in online teamwork. Interviews with children and their parents further revealed key communication channels, benefits, and challenges. Notably, parents played multiple roles in supporting their children's collaboration, especially through modeling and guidance. This study contributes to the design of long-term online collaborative learning interventions for children within computer-supported collaborative learning (CSCL) communities.

## 1. Introduction

The advancement of information and communication technologies has transformed how people work and collaborate. Online platforms facilitate seamless connectivity, fostering a globalized workforce that transcends geographical barriers. As online collaboration becomes the norm in professional settings, governments, educators, and industry leaders emphasize the need for children to develop essential 21$^{st}$-century skills (Larson & Miller, 2011), including digital literacy, communication, collaboration, problem-solving, and creativity. These skills are increasingly crucial for preparing students to navigate future online collaborative environments.

Project-based online collaborative learning provides an authentic context for children to develop collaboration skills, effectively simulating real-world teamwork. It is particularly crucial for upper primary school students (Grades 3–6, ages 8–13), as this stage marks rapid development in social, communication, and problem-solving abilities (Astuti et al., 2022; Goagoses et al., 2022). At this age, students usually have basic digital learning skills, enabling them to engage in online collaborative learning. Additionally, Online project-based learning provokes a deeper level of engagement than traditional education models, stimulating critical thinking, creativity, and problem-solving skills, all within a collaborative setting (Amissah, 2019; Hira & Anderson, 2021). Given its potential to support young learners, online collaborative learning has drawn significant attention from Computer-Supported Collaborative Work (CSCW) and Human-Computer Interaction (HCI) researchers (Li et al., 2023; Mutawa & Sruthi, 2024; Oygür et al., 2020).

However, even for adult learners, the rate of dropout of online learning systems is generally higher than drop-out of a usual campus learning environment (Aydin & Öztürk, 2019; Chuang & Ho, 2016; Zhenghao et al., 2015). According to Simpson's report (Simpson, 2013), the graduation rates of most online higher education are even lower than 10%. For online project-based learning (PBL), particularly due to its nature as a microcosm of real-world collaborative work, the complexity arises from a variety of task types, numerous tools involved, and extensive communication and collaborative skills (Zha et al., 2025, 2024). The wide array of task types necessitates the use of various tools and platforms, potentially leading to cognitive overload and technical difficulties for both students and instructors. In addition, online PBL often involves extensive collaboration, bringing together a broad range of team members (Thomas & MacGregor, 2005). This dynamic can introduce issues of coordination and cooperation, which can be exacerbated in a virtual environment due to communication barriers and lack of face-to-face interaction. Children face several challenges in PBL, including mastering the necessary technology, managing time effectively, dealing with potential delays in support or feedback, staying focused amidst home distractions, coping with feelings of social isolation, accessing reliable internet and necessary equipment, and depending on varying degrees of parental involvement (Aldabbus, 2018; Jerzembek & Murphy, 2013). There are more difficulties due to their limited attention resources and limited experience with online learning tools (Kim, 2020; Sun & Kim, 2023;

CONTACT Jiangtao Gong gongjiangtao2@gmail.com Tsinghua University, Beijing, China

Xue et al., 2025). These factors complicate the already complex learning process and require additional support and resources to ensure success in online PBL. Besides, the majority of existing research related to children's online collaborative learning focuses on the development of specific skills or tasks (Bitter et al., 2016; Sharma et al., 2019; Tsuei, 2011), with little emphasis on fostering cooperation on comprehensive issues within an authentic context.

In response to challenges in supporting children's online collaborative learning, we introduce COLP, a 16-week educational intervention for children aged 8–13. A study with 67 primary school students showed that COLP effectively engaged children in team-based projects. Interviews with students and parents highlighted the benefits of multi-threaded collaboration in enhancing engagement and learning outcomes. Parental involvement was also crucial, as parents actively guided and behaviour collaborative skills. This paper examines these challenges through the COLP framework. Sections 1 and 2 provide an overview of COLP and its five key design principles derived from learning theories. Sections 3 and 4 describe the study methodology, and Section 5 presents key findings. Finally, Sections 6 and 7 discuss design implications and future directions for improving children's online collaborative learning experiences. The main contributions of our work are as follows:

- COLP, a well-designed online collaborative learning program for primary school students using multiple platforms and tools;
- Design guidelines for children's online collaborative learning based on multiple learning theories;
- A long-term, project-based children's online collaborative learning empirical study;
- A rich understanding of communication channels, benefits, and challenges of children's online collaborative learning;
- A focused study of parental involvement during children's online collaborative learning;
- A deep discussion about the design guideline reflections and design implications for an online collaborative learning platform for children.

## 2. Literature review

### 2.1. Online collaborative learning for children

The communities of Computer-Supported Cooperative Work (CSCW) and Computer-Supported Collaborative Learning (CSCL) have long been dedicated to exploring how Information and Communication Technology (ICT) can support collaborative learning (Ludvigsen & Mørch, 2010; Liaqat & Munteanu, 2020; Yadav et al., 2019). Early CSCL research primarily focused on the collaborative learning processes of college students using computer-mediated communication for small- and large-group interactions (Stacey, 1999). In recent years, an increasing number of studies have shifted their attention to online collaborative learning for children, investigating the effects of elementary school students' reading skills, peer interaction, and self-concept in CSCL environments (Bouta & Retalis, 2013; Hourcade et al., 2002; Tsuei, 2011). For instance, Bitter et al. found that regular use of Sokikom, an online collaborative math program, significantly improved elementary students' math achievement and motivation (Bitter et al., 2016). Specifically, students who regularly used Sokikom scored 18% higher on the CAASPP math test and exhibited greater increases in motivation and positive attitudes towards learning mathematics, regardless of their teacher or school. Additionally, online learning systems have been shown to support conversational language learning and digital collaborative storytelling (Sharma et al., 2019).

Nowadays, remote socializing such as learning, communication, and collaboration has become increasingly crucial for students, with some studies specifically discussing the opportunities and challenges of remote collaboration (Chung Galdo et al., 2022; Gui et al., 2021; Yuan et al., 2021). For example, Galdo et al. reported positive feedback regarding students' perceptions of remote collaboration, with some stating that they felt more successful due to working with a partner. This suggests the potential of remote collaboration as a novel learning approach for young learners (Chung Galdo et al., 2022). However, research on synchronous remote collaboration among children remains in its infancy. Current remote collaboration platforms for children often emphasize asynchronous collaboration on long-term projects, such as RALfie (Orwin et al., 2015) and Scratch (Aragon et al., 2009), which primarily focus on fostering creative content rather than facilitating remote socialization and collaboration (Aragon et al., 2009; Orwin et al., 2015). Moreover, most remote collaboration systems, such as Tabletop Teleporter (Odenwald et al., 2020) and Sharetable (Yarosh et al., 2013), are designed to support child-adult collaboration (e.g., between children and parents or teachers), with relatively few systems focusing on child-to-child collaboration. While Angelia et al. proposed a game involving remote collaboration among children, their work lacked a systematic description and specific evaluation of remote collaboration (Angelia et al., 2015).

Although prior studies have explored the impact of online collaborative learning on children, the collaborative tasks they investigated were largely skill-based, such as mathematics, reading, programming, and storytelling. For example, collaborative projects in the Scratch community are child-initiated and rely on community-generated learning materials (Cheng et al., 2022; Dasgupta et al., 2016). However, these tasks do not involve more complex learning models, such as project-based learning, problem-based learning, STEAM programs, or maker programs, which align more closely with authentic and natural collaborative learning scenarios. Furthermore, there is a lack of remote synchronous collaboration systems specifically designed for child-to-child collaboration. To address this gap, this study designs a real-world project using a project-based approach to explore the opportunities for long-term online collaborative learning among children.

### 2.2. Parental involvement in children's learning

Parents play a critical role in children's learning, and numerous studies have explored various aspects of parental engagement and support (Cano-Hila & Argemí-Baldich, 2021; Mendez & Swick, 2018). Drawing on Bronfenbrenner's Ecological Systems Theory, which emphasizes the importance of social interaction and scaffolding within the microsystem of children's development (Darling, 2007; Li et al., 2023; Seginer, 2006), research has shown that parental roles in learning are influenced by factors such as socioeconomic status (SES) and children's developmental stages. For example, Yu et al. identified three parental roles—spectator, scaffolder, and teacher—in the learning processes of children aged 3–9 years (Yu et al., 2020), while Barron et al. explored parental involvement in middle school students' development of technological fluency, identifying seven distinct roles, including teacher, collaborator, and resource provider (Barron et al., 2009). These findings highlight how parental involvement shifts as children grow older, moving from direct guidance for younger children to more indirect support, such as fostering independence and providing emotional encouragement, for older children. With the rise of technology-based education, this evolution in parental roles becomes even more pronounced, particularly in the context of online collaborative learning. This shift underscores the need to investigate how parents adapt their roles to effectively support children's learning in digital environments.

With the widespread adoption of online learning, parents face increasing challenges that demand significant time, energy, and adaptability (Dong et al., 2020; Wang et al., 2024). In technology-based learning contexts, parents often take on multiple roles, such as designing learning activities, sourcing resources, managing progress, and even directly teaching their children (Liu et al., 2025; Yu et al., 2021). These roles are further influenced by children's age and developmental needs. For younger children, parents may need to provide constant supervision and hands-on guidance, while for older children (typically aged 8–13), the focus often shifts to fostering self-regulation, encouraging collaborative problem-solving, and supporting the development of digital literacy skills (Gelir & Duzen, 2022; Han et al., 2022). For instance, Mills et al. noted that parents of older children often act as facilitators, helping their children navigate complex tasks and mediate group interactions, rather than directly teaching content (Mills et al., 2021). Additionally, socioeconomic factors play a significant role in shaping parental involvement; families with lower SES may struggle to provide adequate technological resources, while higher SES families are better equipped to offer both educational and emotional support (Hohlfeld et al., 2010; Yardi & Bruckman, 2012). These dynamics highlight the importance of tailoring parental support to the specific developmental and contextual needs of children in online learning contexts.

In collaborative learning settings, particularly for children aged 8–13, parents play a crucial role in promoting group discussions, providing resources, offering technical support, and mediating group dynamics. Research by Druga et al. demonstrated that parents can enhance children's collaborative AI learning by facilitating peer learning, organizing family challenges, and guiding group activities, which not only improve teamwork and communication skills but also deepen children's understanding of complex concepts (Druga et al., 2022). However, as children in this age group are developing greater independence, parents need to strike a balance between providing support and allowing autonomy. Over-involvement may hinder children's ability to self-regulate and collaborate effectively, while under-involvement may lead to disengagement or frustration. Therefore, for parents of 8–13-year-old children, key considerations include fostering a supportive learning environment, encouraging open communication, and gradually reducing direct supervision as children become more confident in their abilities. This study aims to further explore how parents can effectively support project-based online collaborative learning for this age group, providing insights to inform both platform design and family education policies.

### 2.3. Design considerations for children's long-term online collaborative learning

Drawing upon the combined pedagogical experiences of our interdisciplinary team, composed of three educators and three educational researchers, we have identified several challenges to successful online collaborative learning for children. These challenges include low retention rates, hard-to-maintain students' attention, inadequate real-time guidance and feedback, underdeveloped communication and collaborative skills, and the lack of child-friendly interfaces in online collaborative platforms. In this regard, our research team has embarked on a thorough exploration of relevant theories, including developmental psychology, constructivism, and social learning theory, to underpin the design of our Children's Long-term Online Collaborative Learning. After several rounds of discussions and iterative refinement, we converged on a set of five design guidelines based on five relevant theories as follows.

#### 2.3.1. Active learning

Active learning is a method of learning in which students are actively or experientially involved in the learning process and there are different levels of active learning, depending on student involvement. Bonwell and Eison (1991) state that students participate in active learning when they are doing something besides passively listening. According to Hanson and Moser (2003) using active teaching techniques in the classroom creates better academic outcomes for students. Scheyvens et al. further noted that by utilizing learning strategies that can include small-group work, role-play and simulations, data collection, and analysis, active learning is purported to increase student interest and motivation and to build students' critical thinking, problem-solving, and social skills (Scheyvens et al., 2008).

There is a wide range of alternatives for the term active learning, such as learning through play, technology-based learning, activity-based learning, group work, project methods, etc. The common factors in these are some significant qualities and characteristics of active learning. Active learning is the opposite of passive learning. It is learner-centered, not teacher-centered, and requires more than just listening; the active participation of students is a necessary aspect of active learning. In active learning, children are encouraged to come up with creative solutions for projects. This fosters creativity and innovation, which are vital skills in the 21st century (Lamb et al., 2017; Luna Scott, 2015). Furthermore, online project-based learning often involves teamwork. Active learning promotes collaboration as children need to work together, share ideas, and cooperate to complete projects. This enhances their social skills and helps them learn to work effectively in teams. Based on the above, we developed our design guideline 1:

> Design Guideline 1: Design dynamic and interactive activities to enhance students' active learning.

### 2.3.2. Observational learning

Observational learning is learning that occurs through observing the behaviour of others, as children observe their peers, they also learn the importance of teamwork, collaboration, and the sharing of ideas. They get to understand different perspectives and approaches to solving a problem or completing a project, which can enhance their own problem-solving skills, which is important for collaborative learning. In humans, this form of learning seems to not need reinforcement to occur, but instead, requires a social model such as a parent, friend, or teacher with the surroundings. Particularly in childhood, a model is someone of authority or higher status in an environment. In animals, observational learning is often based on classical conditioning, in which an instinctive behaviour is elicited by observing the behaviour of another, but other processes may be involved as well (Shettleworth, 2009). Besides, Online learning requires certain behaviors and skills, such as digital literacy, time management, and self-discipline. By observing their teachers and peers, students can learn and adapt these behaviors for their own learning. It is a form of social learning that usually takes an in-person form. We believe that observational learning will also benefit online collaborative learning for children. Thus, we formulated our design guideline 2:

> Design Guideline 2: Build an online collaborative environment to support online observational learning and facilitate communication among students.

### 2.3.3. Parental involvement

In an online learning environment, parental involvement is often necessary to ensure that children stay on track with their projects. Parents can help create a conducive learning environment at home, supervise their children's online activities to ensure safety, and guide them when they encounter problems. Parents' involvement in their children's education and parental warmth has been linked to many positive child outcomes (Ogg & Anthony, 2020). Educators consider parental involvement an important ingredient in the remedy for many problems in education. Parents can enhance their children's learning by providing additional resources, discussing concepts, or even learning together with their children. They can also help their children make connections between what they are learning and real-world applications. It is important to set a role for parents to be involved in their children's learning. Parents also need to provide technical support to help children overcome technological barriers as they use a variety of online learning platforms (Anastasiades et al., 2008; Cumbo et al., 2021). Thus, we formulated our design guideline 3:

> Design Guideline 3: Create different roles to involve parents in students' learning and form a learning community

### 2.3.4. Scaffolding

Scaffolding is a pedagogical approach that involves providing learners with the necessary support to enable them to accomplish tasks and develop understandings that are beyond their immediate grasp (Hammond & Gibbons, 2005). Its theoretical underpinnings are rooted in Vygotsky's socio-cultural theory (Hammond & Gibbons, 2005), which underscores the importance of social interaction and support in learning (Pablo Hourcade, 2008). The facilitator plays a pivotal role in shaping the learners' experience, with the ability to modulate the pace of the activity, influence group dynamics, and guide learners towards the desired outcomes. Scaffolding is particularly crucial in educational settings, where teachers or more advanced peers assist learners by progressively tapering off support as the learner becomes more proficient.

In the realm of online project-based learning for children, scaffolding is indispensable for fostering comprehension and tailoring the learning experience to meet individual needs. Various forms of scaffolding can be implemented, including but not limited to task behaviour, provision of step-by-step instructions, feedback delivery, posing guiding questions, and employing digital tools to navigate learners through tasks. These strategies collectively enhance the impact of online project-based learning on children. Consequently, we have revised our fourth design guideline to reflect the importance of scaffolding in online project-based learning:

> Design Guideline 4: Implement scaffolding strategies to foster students' deep learning

### 2.3.5. Gamification

Gamification is a strategic approach aimed at enhancing systems, services, organizations, and activities by creating experiences similar to those found in games, with the goal of motivating and engaging users (Koivisto & Hamari, 2019). In the context of education, the gamification of learning leverages video game design principles and game elements to create engaging learning environments (Shatz, 2015; Sailer et al., 2017). This approach seeks to maximize enjoyment and engagement by capturing learners' interest and inspiring them to persist in their learning journey (Hsin-Yuan Huang & Soman, 2013).

In the specific context of children's online project-based learning, gamified incentive mechanisms can play a pivotal role in stimulating student engagement. These mechanisms tap into children's natural playfulness and competitive spirit, transforming learning into a more enjoyable and interactive process. For example, features such as leaderboards provide children with timely feedback on their progress, allowing them to track their achievements and sustain motivation throughout long-term PBL projects. Moreover, gamified systems not only encourage active participation but also foster a sense of accomplishment and collaboration. By integrating incentives into COLP, children are motivated to engage more deeply with their peers, facilitating peer learning and teamwork. These mechanisms create an environment where children can learn through interaction, competition, and cooperation, which are critical for the success of project-based learning. To address the unique challenges of COPL and harness the benefits of gamification, we developed Design Guideline 5:

> Design Guideline 5: Cultivate motivation and engagement through an incentive-based system

## 3. Designing COLP program

The COLP program is meticulously crafted to enhance the online collaborative learning experience for children, integrating various learning activities, tasks, environments, and management practices. Building upon the insights gathered from Section 2.4 and the associated review of relevant theories, we have derived five design guidelines aimed at optimizing the program for children's collaborative learning in a virtual setting and addressing their unique needs. In this section, we delve into the application of these guidelines within our program.

Our learning project is rooted in the principles of project-based learning (PBL) and is tailored for children aged 8–13 who behavior a keen interest in technology, creativity, and gaming. Participants are encouraged to form teams with their peers and embark on an adventurous journey, working together on an entrepreneurial programming game project. It is important to note that the participants were strangers to each other at the onset of the program.

In this project, participants not only gain proficiency in Scratch programming, digital art, and music production but, more importantly, they develop the ability to transform their ideas into products and exhibit creative problem-solving skills within complex environments.

### *3.1. Design guideline 1: Design dynamic and interactive activities to enhance students' active learning*

To acquire a profound comprehension of the experiences and challenges inherent in children's online collaborative learning, we devised a comprehensive 16-week STEAM-focused online collaborative learning program aimed at elementary school students aged 8–13. It included learning three types of skills: Scratch programming, digital art creation, and digital music creation.

In orchestrating the learning project, we meticulously selected a triad of diverse tasks, meticulously aligning them with the students' interests and prior experiences, while ensuring their accessibility for independent completion. This approach was integral to facilitating the seamless acquisition of varied skills. The tools employed—Scratch for programming, Piskel for digital art, and GarageBand for music creation—were chosen for their educational efficacy and user-friendliness. With the support of guided curriculum, students were equipped to adeptly navigate and utilize these tools, as illustrated in Figure 1. The COLP program's activities were designed to foster active engagement, encompassing team-building exercises, skill acquisition sessions, and collaborative endeavors on the final project.

To nurture the students' collaborative problem-solving skills, we encouraged each team to take on the challenge of creating their own programming game. This process was enhanced by introducing the concept of online team-building, where students were guided on how to collaborate effectively with their team members in a virtual environment. This method fostered a sense of equality and mutual dependence, as every student was seen as a vital contributor to the team's success. This innovative educational approach not only enriched the students' active learning experience but also served as a valuable foundation for developing essential collaborative problem-solving skills, preparing them for future endeavors in both academic and professional settings.

### *3.2. Design guideline 2: Build an online collaborative environment to support online observational learning and facilitate communication among students*

The online collaborative environment is significant for children's learning. In our program, teachers, students, and parents primarily utilize four collaboration application types in online collaborative learning: the Xiaoe-tech learning platform; the WeChat instant messaging application; the Voov Meeting video conferencing application; and creative programming tools Scratch, Piskel, and GarageBand. Xiaoe-tech is mostly used for class activities, lesson releases, and other tasks. During the program, the teacher taught by releasing recorded lessons on Xiaoe-tech. Students were given relevant assignments at the same time to help them remember what they had learned. The teachers' recorded classes were available for students to view on Xiaoe-tech. Students were asked to complete both individual and team assignments posted on Xiaoe-tech after watching lessons. Students were to upload a screenshot of their finished assignments to Xiaoe-tech upon completion. All the other students and their parents could view the screenshot on Xiaoe-tech, and some would comment on and like other students' work. As illustrated in Figure 2, a children's online collaborative learning environment is depicted. Children use iPads to access recorded and live courses. They practice programming, art, and music creation on computers. Additionally, they use mobile phones to interact and collaborate with peers, facilitating discussions and group activities.

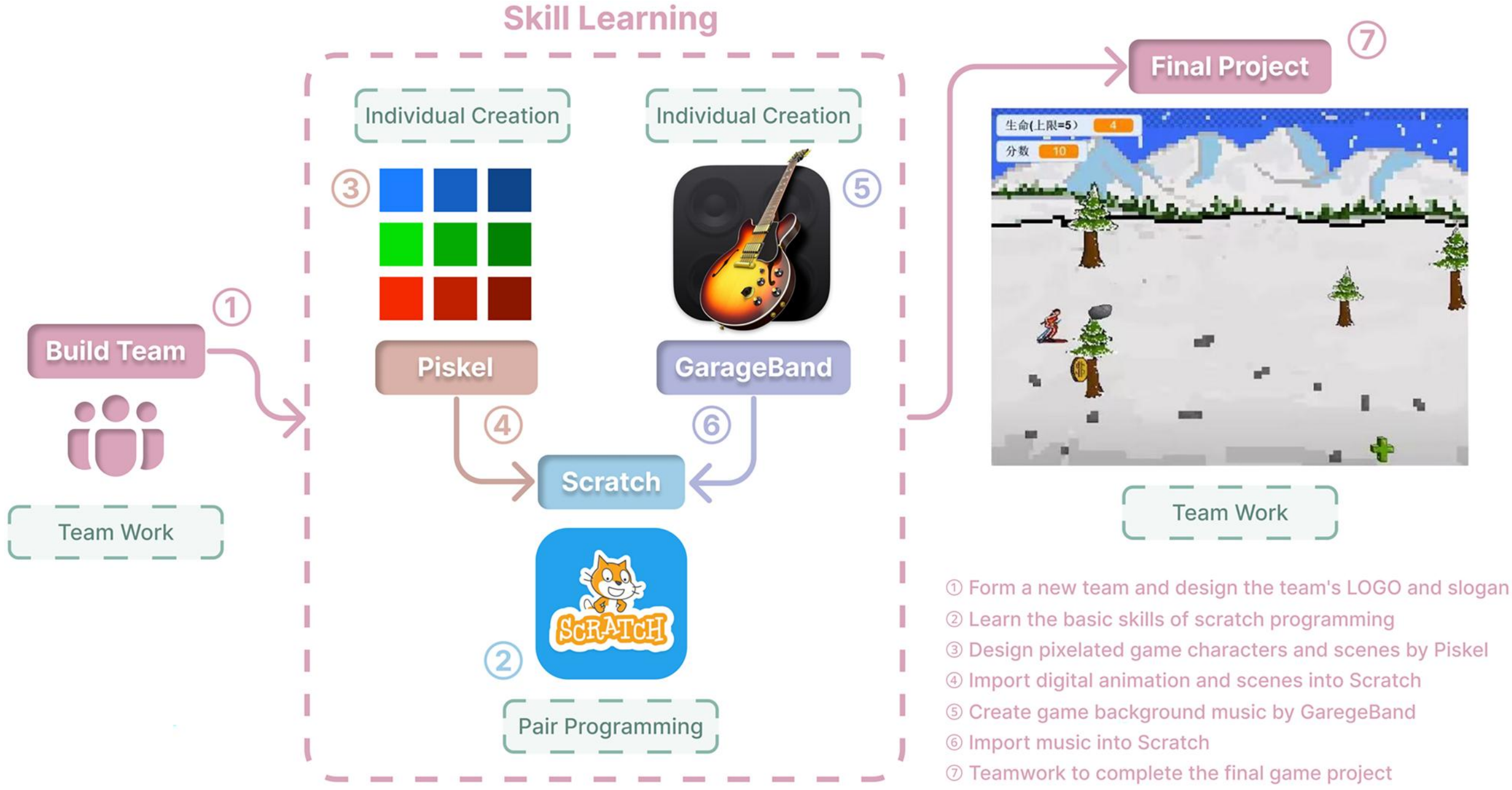

**Figure 1.** COLP program's activities design.

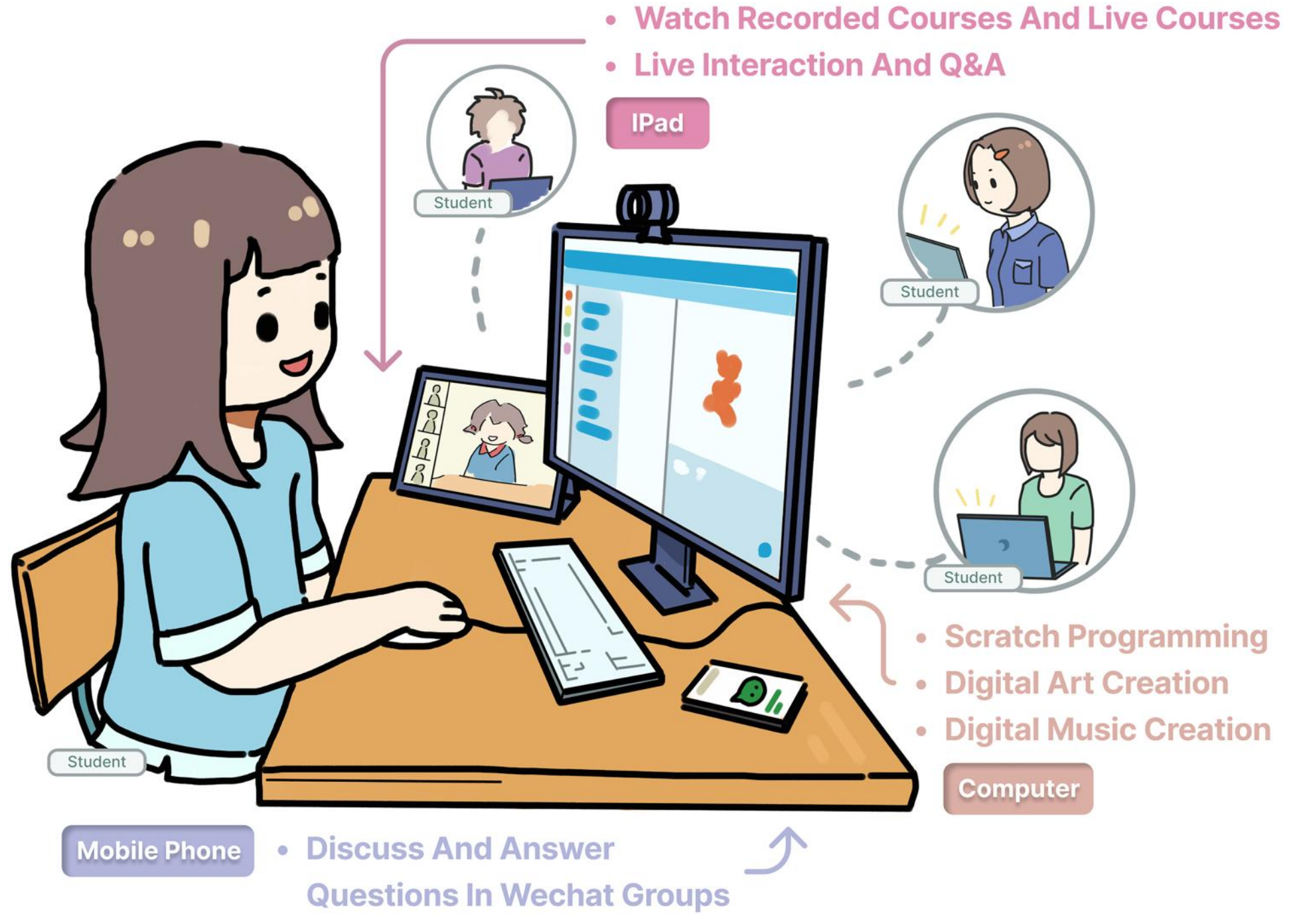

**Figure 2.** Children's online collaborative learning environment design.

### *3.3. Design guideline 3: Create different roles to involve parents in students' learning and form a learning community*

Parents play a very important role in children's learning, especially in the online collaborative learning environment. To seamlessly integrate parents into their children's learning pathway, we have crafted explicit roles and corresponding tasks, meticulously designed to enhance parental participation while maintaining the integrity of the learning process.

Firstly, parents assume the role of "Product Publicity Officers," where they are entrusted with "publicity tasks." These tasks involve providing affirmative reinforcement to their children and celebrating their achievements through public acknowledgment and sharing. This role is vital as it not only boosts the children's morale but also cultivates a nurturing and supportive learning environment. In addition, parents serve as "Program Testers," a role that equips them with "test tasks" and program test sheets, guiding them through a structured process of evaluating the game programs devised by their children. This facilitates the provision of constructive feedback and ensures an active parental engagement in the learning journey of their children, without crossing the line into completing assignments or programming tasks on their children's behalf. The roles of "Product Publicity Officer" and "Program Tester" were meticulously chosen to ensure that parents do not inadvertently find themselves completing programming tasks or other assignments for their children. Instead, these roles are constructed to engage parents in a supportive and supervisory capacity, contributing constructively to their children's educational progression.

By fostering a learning community that brings together parents, children, and teachers, we are establishing a collaborative network that significantly enhances parental involvement. This inclusive and community-driven approach ensures a comprehensive support system, empowering students to excel in their online collaborative learning endeavors while ensuring that parents contribute positively and constructively, thereby enhancing the overall learning experience.

### 3.4. Design guideline 4: Implement scaffolding strategies to foster students' deep learning

As illustrated in Figure 3, this project-based online collaborative learning program leverages a blend of live streaming and video-recorded content. Using Xiaoe-tech, a digital learning management platform, we disseminate 15–20 min instructional videos accompanied by relevant learning materials every week. Additionally, live sessions via Voov Meeting, a video conferencing tool, are held every 2–3 weeks primarily for real-time Q & A. The program is bookended with a live opening ceremony and a concluding debriefing session.

Post each recorded session, detailed learning assignments are posted to engage the students further. To foster deep learning and encourage active participation, we have curated a variety of learning tasks, both at an individual and team level. Students are prompted to engage in collaborative activities, such as jointly creating team names and logos, and partaking in peer review for program tests during team tasks. On the other hand, individual tasks challenge students to hone their skills in game programming, digital art, and music creation.

This structured scaffolding approach ensures a balanced blend of collaborative and individual work, aiming to enhance students' competencies and foster a collaborative learning environment. Through these meticulously designed tasks, we aim to facilitate a deepened understanding and mastery of the skills at hand, aligning with our objective of scaffolding students' learning journey in the online realm.

### 3.5. Design guideline 5: Cultivate motivation and engagement through an incentive-based system

In our online PBL program, we have integrated an incentive-based system inspired by gamification principles to encourage and recognize students' achievements and participation. This approach introduces a new layer of interaction and engagement, fostering motivation for long-term involvement in online PBL activities. Upon attempting an assignment tied to our incentive system, students encounter a series of challenges, each accompanied by detailed information, including the potential rewards and recognition for successful completion. This clarity is aimed at motivating students, and providing a clear path toward achieving their goals.

To bolster the competitive and collaborative spirit of the program, we have incorporated a leaderboard feature within our digital environment. This feature serves dual purposes, evaluating both individual and team tasks to accommodate the varied needs of our student cohort. Students are required to complete all assigned tasks within the given timeframe to ascend the leaderboard rankings and gain recognition for their efforts. Real-time rankings, diligently updated by our teaching assistants based on weekly performance, ensure an accurate and current reflection of each student's engagement and progress. Individual rankings highlight student participation, while team rankings capture the collective activism of team members and their progress in team tasks.

By integrating gamification elements into our PBL approach, this incentive-based system serves to motivate task completion, while also contributing to a dynamic and engaging learning experience. The ultimate goal is to instill a sense of achievement and motivation among students, sustaining their engagement in online PBL activities over the long term.

## 4. Methodology

In our study, in order to deeply understand the experience and challenges of children's online collaborative learning, we conducted a long-term study with a COLP program involving primary students in more than five provinces in China. During the study, we used a mixed method, which included observations and semi-structured interviews with children and their parents. This study received IRB approval from our institution.

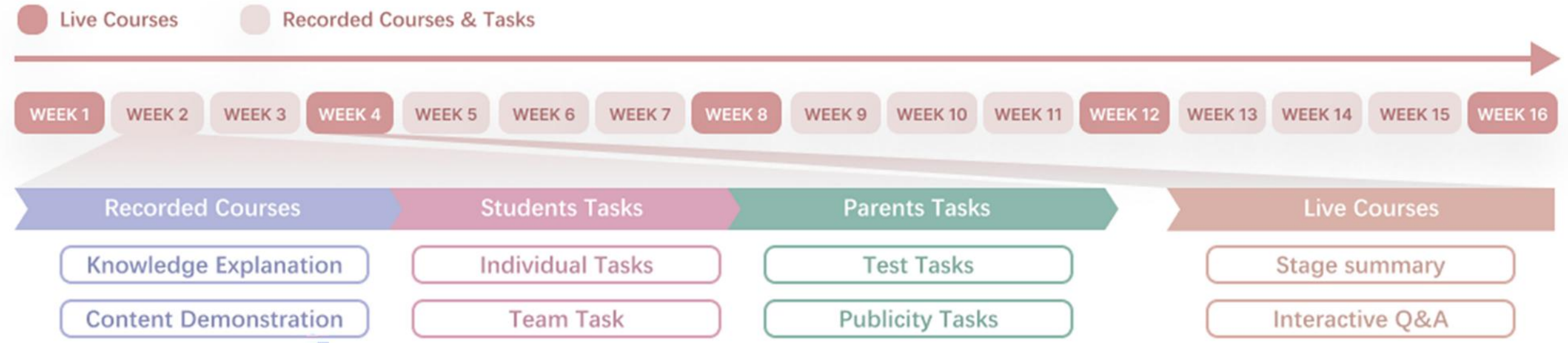

Figure 3. Online collaborative learning program.

### 4.1. Research context

China's education system actively integrates technology into learning, particularly for upper primary students (grades 4–6), who are introduced to digital tools as part of their curriculum. The government strongly promotes the use of technology in education, as reflected in national policies encouraging digital literacy and online learning platforms. The Education Informatization 2.0 Action Plan, released in 2018, marks a significant milestone in China's digital education strategy (Yan & Yang, 2021). This policy emphasizes innovation-driven development and the deep integration of information technology into education. These policies encourage the exploration and application of emerging technologies across various educational contexts.

The adoption of online learning was further accelerated by the COVID-19 pandemic, during which both formal and informal learning transitioned to digital platforms. In early 2020, the Ministry of Education of China mandated that all schools and universities shift to online instruction, ensuring educational continuity during lockdowns (Dong et al., 2020). Educators and curriculum developers actively introduced digital learning resources and educational apps to support children's home learning (Chen et al., 2020). Many parents, following recommendations from teachers and educational authorities, facilitated their children's participation in online learning (Lau & Lee, 2021). This large-scale shift reinforced students' familiarity with online learning environments and normalized the use of technology for education, even among younger children. Beyond formal schooling, online education plays a significant role in informal and after-school learning (Zheng et al., 2019). Many Chinese students participate in extracurricular programs, particularly in STEAM fields such as mathematics, programming, and robotics, where a substantial portion of instruction is delivered online. This reflects a broader societal trend in which parents seek supplementary education opportunities for their children, often leveraging digital platforms.

Parental involvement in children's education is a deeply embedded cultural norm in China. Parents not only supervise schoolwork but also facilitate access to online learning resources, often possessing sufficient digital literacy to support their children's education (Gong et al., 2021). This strong parental engagement, combined with students' early exposure to digital learning in extracurricular settings, creates a supportive environment for online collaborative learning, making interventions like COLP particularly relevant in the Chinese context.

### 4.2. Participants

In our study, a total of 67 children aged 8 to 13 years old participated (Mean age = 8.87, SD = 0.85). We reached out to potential participants by sharing an electronic poster with various parent groups and by posting it on WeChat. The poster provided an overview of the program, describing it as a STEAM-type children's collaborative learning experience delivered online. Additionally, it included details about the program's schedule.

To facilitate the registration process, we embedded a QR code on the poster, linking directly to an online questionnaire. Interested parents were required to complete this questionnaire as a preliminary step for their child's participation. The questionnaire was divided into two main sections. The first section collected demographic information about the participating children. During this stage, we identified and excluded some children who did not fall within the 8–13 age range. The second section of the questionnaire gathered information pertinent to the experiment. This included inquiries about the children's interest and familiarity with programming, their experience with digital games, and an assessment of their collaborative skills. To ensure ethical compliance and prioritize the well-being of the participating children, we obtained written informed consent from the parents and oral assent from the children themselves.

To enhance the effectiveness of the project's implementation, we randomly recruited three experienced teachers to jointly design the COLP program. Additionally, we also recruited five teaching assistants with educational backgrounds to participate in our project. The inclusion of these professionals provided robust support for the smooth progress of the project, ensuring both the quality of the learning content and the efficacy of the learning process.

### 4.3. Procedure

During the course of our online collaborative learning program, we implemented a mixed-method study that incorporated learning data analysis, observation of all 67 participants, and semi-structured interviews with a subset of 46 participants (23 parents and 23 children). The aim was to gain a comprehensive understanding of the children's experiences in online collaborative learning. To facilitate observation of the collaborative learning process and provide support for student discussions, we established a WeChat group that included all 67 students, their parents, and teacher assistants. In addition to this main group, we created 17 smaller, official WeChat groups. These groups were organized based on the students' ages and genders, each consisting of 3–4 students, their parents, and teacher assistants.

Prior to the start of the program, we secured consent from all the teachers and parents of the 67 participating children. This allowed us to monitor and record activities across all groups throughout the duration of the course project. The types of data collected included transcripts of Q & A discussions, photos, and videos from WeChat Group. Upon the completion of the learning program, we randomly conducted interviews with some of the parents and students.

These interviews were crucial for providing additional context to the observational and learning data, helping us to better understand the intricacies of the online collaborative learning experience from both the students' and parents' perspectives.

### 4.4. Learning data and observation

In this study, the first author acted as a program teaching assistant and collected rich course-related learning data. Firstly, we observed the teachers' process of recording and uploading course videos and evaluating students' task performance. Secondly, the online learning platform recorded rich student operational data, including the number of students who watched recorded courses and submitted assignment tasks (which provides quantitative evidence for an in-depth understanding of the student learning process).

Because some students did not have their own accounts, they would share accounts with their parents to interact with other students or teachers. Thus, we observed rich interaction data through WeChat during collaborative learning, which included teacher-student (parent) interactions, student-student (parent) interactions during live streaming and Q & A in WeChat groups, and students sharing their tasks and interactive feedback in the group. In addition, we noted the collaboration and interaction between each group of students (parents) in the small groups and estimated the frequency of such interactions. The data are shown in Table 1.

### 4.5. Interview

At the end of the program, we conducted extensive interviews with children (N = 23) and their parents (N = 23), randomly selected from among all the participants to gain insight into their experience of the online collaborative learning program as a whole. The children comprised 17 boys and 6 girls, with an age range of 8 to 11 (the demographics information see in Supplementary Appendix B1). Due to the technical nature of the course content, we recorded whether students had a relevant knowledge base in programming, etc., before participating in the program. We interviewed students and their parents via video conferencing software. All interviews were audio recorded. Each interview lasted about 30 min to an hour (Alshenqeeti, 2014; Lune & Berg, 2017). The key questions (see Supplementary Appendix B2 for details) were designed as a checklist to help cover all study themes via semi-structured interviews. We first conducted individual interviews with both parents and children to gather their overall opinions regarding the online collaborative course project. Our aim was to understand how children interacted with their team members during the program, whether they sought help from their parents, and whether parents provided assistance. During the interviews, we delved into specific details by asking follow-up questions based on the responses provided by the students and parents. For instance, when a student mentioned that their parents supported their learning, we sought to gain deeper insights by asking questions such as, "In general, how do your parents support you?"; "Do you remember any specific instances when their support was particularly effective?"; and "How does it help you when your mom and dad support you with your learning?" This approach allowed us to explore the authentic experiences and realities of children's online collaborative learning in a more comprehensive manner. All interviews were fully transcribed and reviewed. An inductive thematic analysis approach was applied when behaviour the data, mainly following a step-by-step approach guided by Maguire and Delahunt (2017). First, three researchers conducted a line-by-line open coding of the 46 transcripts. All codes were reviewed together, and agreements were achieved for each. Next, as suggested by Maguire and Delahunt (2017), codes were reviewed and divided into themes and sub-themes to answer our research questions.

**Table 1.** Learning data summary.

| Data source | Data type |
| --- | --- |
| Learning platform (Xiaoe-tech) | Task completion status<br>Learning duration |
| IM group (WeChat) | Student group communication<br>Dialogue data |
| Interviews | Student and parent interviews |

## 5. Results

This section contains our findings in five parts: (1) An introduction to current practices in our children's online collaborative learning program; (2) The communication channels feature of children's online collaborative learning; (3) The benefits of children's online collaborative learning; (4) The challenges of children's online collaborative learning; (5) Parental involvement and guidance in children's online collaborative learning.

### 5.1. Overall feedback

In this subsection, we present an overview of students' engagement and task completion throughout the COLP program. We behavior student retention trends, task participation rates, and key factors that influence sustained participation. Additionally, we highlight both the benefits and challenges observed in collaborative learning experiences of students.

#### 5.1.1. Student retention and task engagement

In this program, we behaviour 67 students' (45 boys and 22 girls) learning data. All are primary school students in Grades 1–5, aged between 8 and 13. Figure 4(a) shows the program's student retention rate, which is calculated by dividing the number of students who completed each course by the total number of students (67) who signed up. As shown in Figure 4(a), as time progressed, the student retention rate trended downward. The retention rate was highest at the beginning of the program (during the first two weeks), with 100% of students attending the course during weeks one and two. A few students began to miss classes from week three. Week five saw a faster decline in the number of students attending the course. By the time the course ended, retention was at its lowest, with only 36% of students attending. More students gradually dropped out as the program progressed. However, the number of participants changed drastically during a two-week period when new course content was introduced. For example, in Figure 4(a), we can observe that compared with week five, the week six retention rate dropped more quickly than it had previously. This may be due to an increase in course difficulty. The first five weeks' content mostly comprises introductions to new software and basic programming concepts. The course content becomes more challenging in the sixth and subsequent weeks, containing complex programming topic instruction, and certain learning challenges.

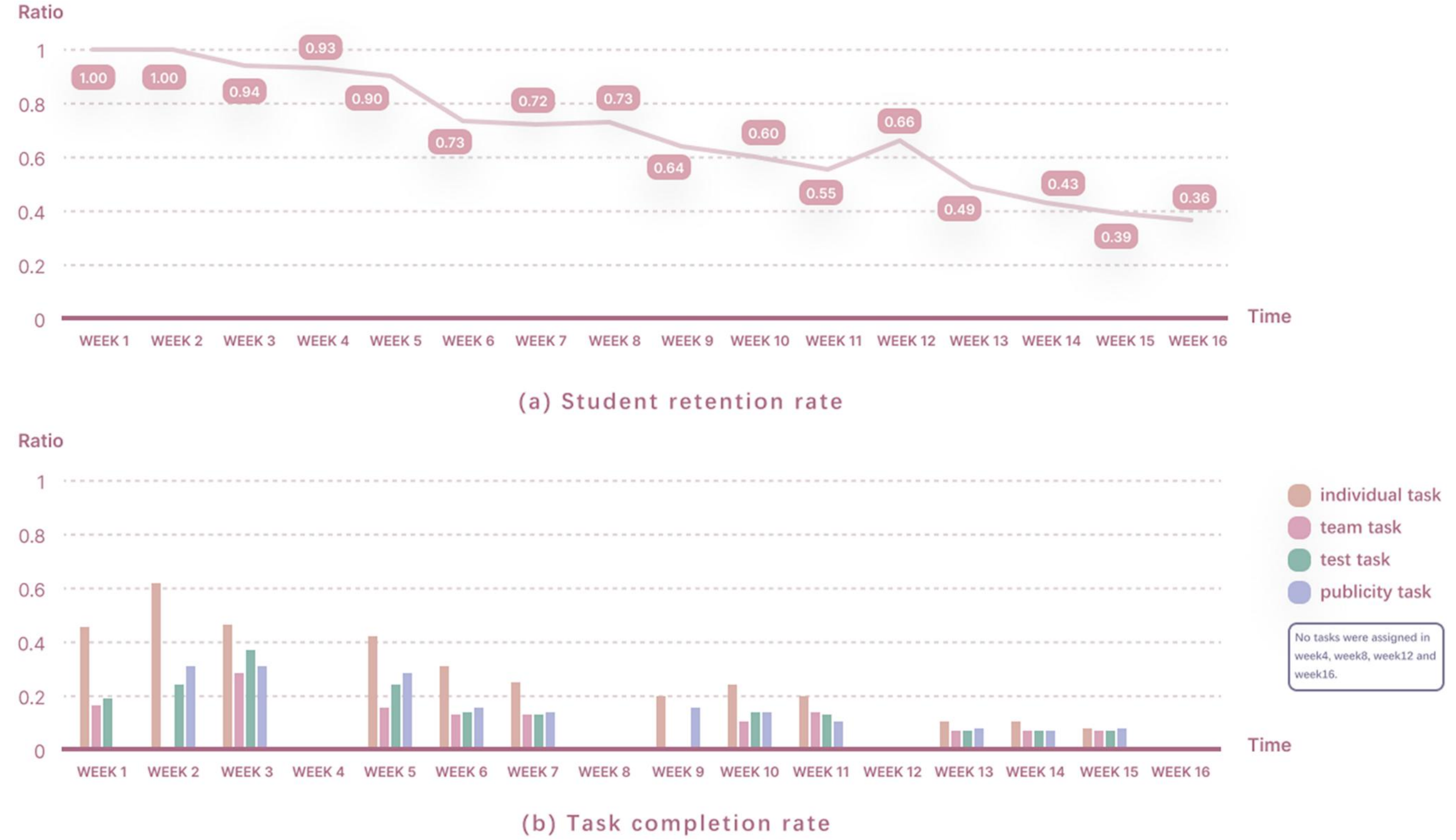

**Figure 4.** Student retention rate and task engagement. (a) Student retention rate; (b) task completion rate.

Following the course, students needed to finish both the individual and team tasks listed on Xiaoe-tech (see Figure 5). After completing these tasks, they would upload screenshots of their finished work to Xiaoe-tech. All of the other participating students and their parents could view all of the uploaded work, and some of them would comment on and like other students' work.

The program tasks – including programming, art, and music composition – all center on creating video games. Students use Scratch, Piskel, and GarageBand to program, create artwork, and create music at various stages of the game's creation. They use WeChat to send screenshots and source code written during the creation process to other team members. Students can capture screenshots and upload them to Xiaoe-tech. During live lessons, their work was displayed and discussed.

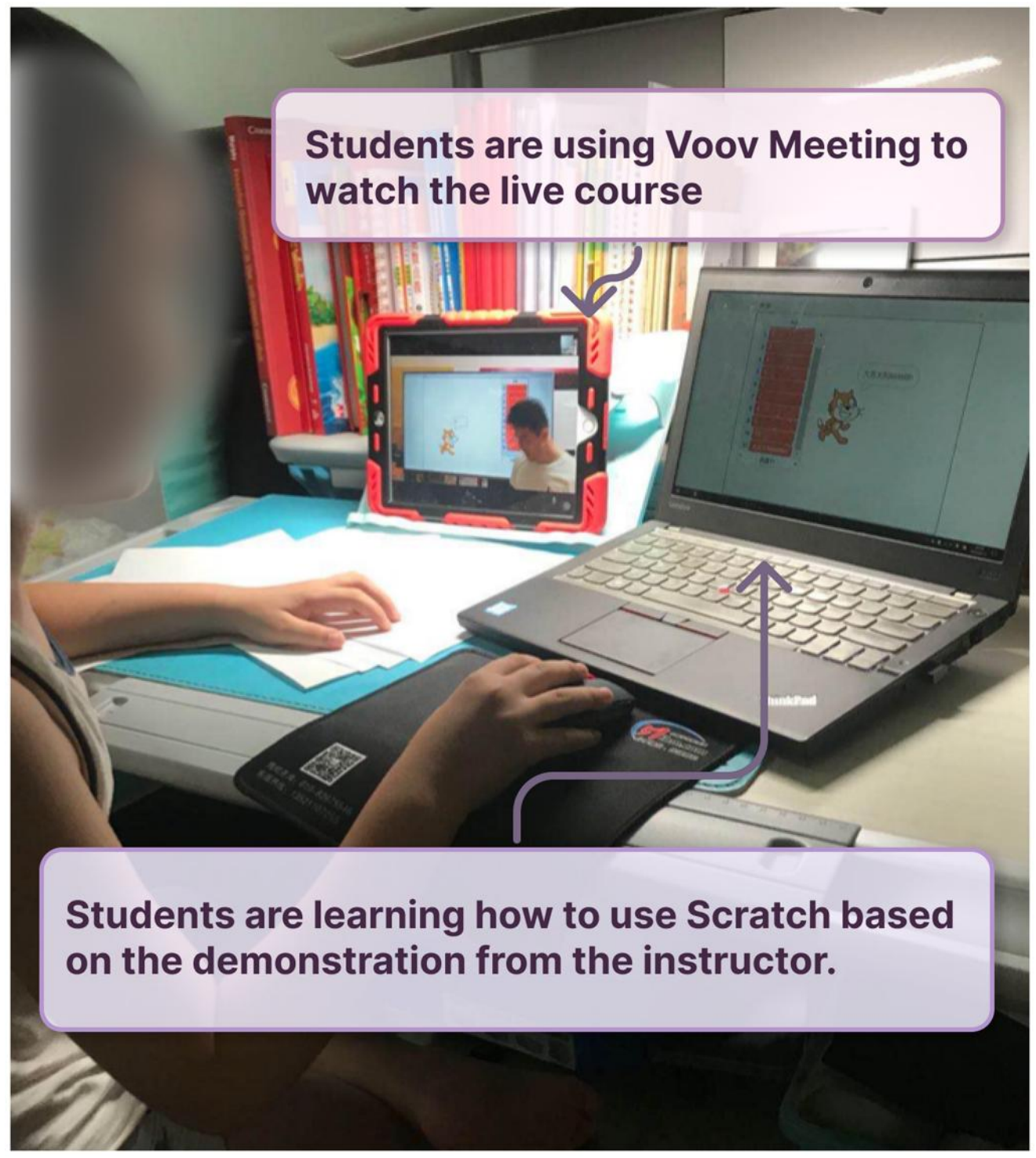

**Figure 5.** The child is watching a live course on Voov Meeting and practicing using scratch.

The online collaborative learning program for children was designed with a series of tasks, such as individual and team tasks for students, and test and publicity tasks for parents. No tasks were assigned during weeks four, eight, 12, or 16. There were no publicity tasks during week one.

During week two, no team tasks were assigned. No team or test tasks were assigned during week nine. As shown in Figure 4(b), completion rates for all four tasks also tend to decrease, with higher completion rates in the first three weeks, and the lowest completion rates in the last three.

The completion rates for various tasks varied significantly in our study. For individual tasks, the highest completion rate was 61%, while the lowest was only 7%. As for team tasks, 27% was the highest completion rate, whereas the lowest was 6%. When it came to test tasks, the highest and lowest completion rates were 36% and 6% respectively. Finally, for publicity tasks, the completion rates ranged from 30% at the highest to 7% at the lowest.

It's worth noting that out of all the 17 participating groups, seven were able to successfully complete the final project. To provide a glimpse into the quality of work produced, we have chosen a representative sample of these successful projects and presented it in Figure 6.

The observed trend in task completion rates aligns with that of retention rates, showcasing a decrease over time. The data reveals that students were generally more successful in completing individual tasks compared to team tasks, highlighting the challenges they faced in collaborative online settings. A potential explanation for this disparity is the inherent difficulty in initiating team tasks, especially when students are not previously acquainted. P1 shed light on this issue, stating, "Online, students did not know each other and there was a lack of team cohesion. Only a small number of students in each group were very active, and most of them did not move to initiate the task of making the game." This points towards the necessity for strategies that foster team cohesion and active participation in online collaborative learning environments.

During the first seven weeks, the individual task completion rate was much higher than the team task completion rate. However, from weeks 13 to 15, the individual task completion rate was close to that of team tasks. It can be found that the variance in students' individual task completion rates is larger than the variance in team task completion rates. During the final weeks, when overall completion rates were lowest, the same group of students completed both the individual and team tasks. This may be because, in good teamwork, group members support each other and motivate them to complete tasks.

Task completion rates are also affected by task type. For example, in week two, the individual task completion rate increased significantly compared with week one; in fact, it was the highest in the whole course. The reason may have been that the second week's task was to "develop the team's name, logo, and slogan." The teacher provided clear and detailed task objectives, and the task type was relatively simple, in line with primary school students' cognitive levels. The parental task completion rate also decreased as the program progressed. However, the parents' publicity task completion rate was higher than that of the more complex test task, which required more time and effort. The completion rate for all tasks showed a relatively consistent trend and tended to be more similar during weeks 13 to 15.

### 5.1.2. Impact of design guidelines on children's online collaborative learning

The design guidelines we implemented in the project have yielded several benefits. However, we have also encountered some challenges. In this section, we summarize them separately. We have designed numerous learner-centered learning activities with the intention of motivating learners' engagement in the learning process. Through our project, we discovered that the course activity design indeed stimulated learners' interest in learning, provided them with a sense of achievement, and encouraged them to actively seek assistance in their learning, thus fostering a proactive learning attitude. As mentioned by P5,

> "The course cultivates children's self-directed learning methods and sparks their interest, enabling them to independently complete course tasks, and my child has been very engaged. After the course, they were able to create games, record them, and share them with other classmates. It was quite fulfilling and taught them how to seek assistance and share with each other."

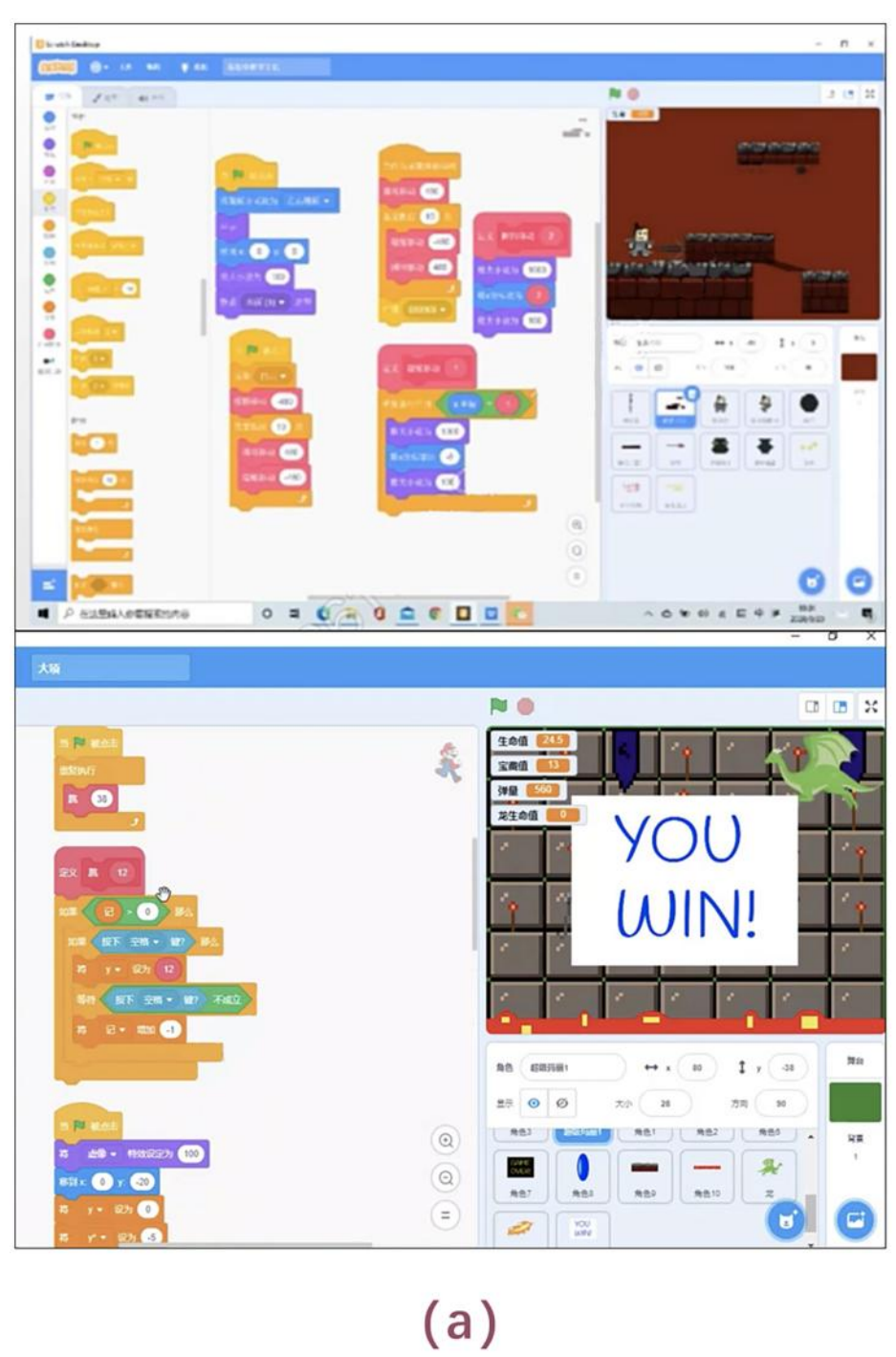

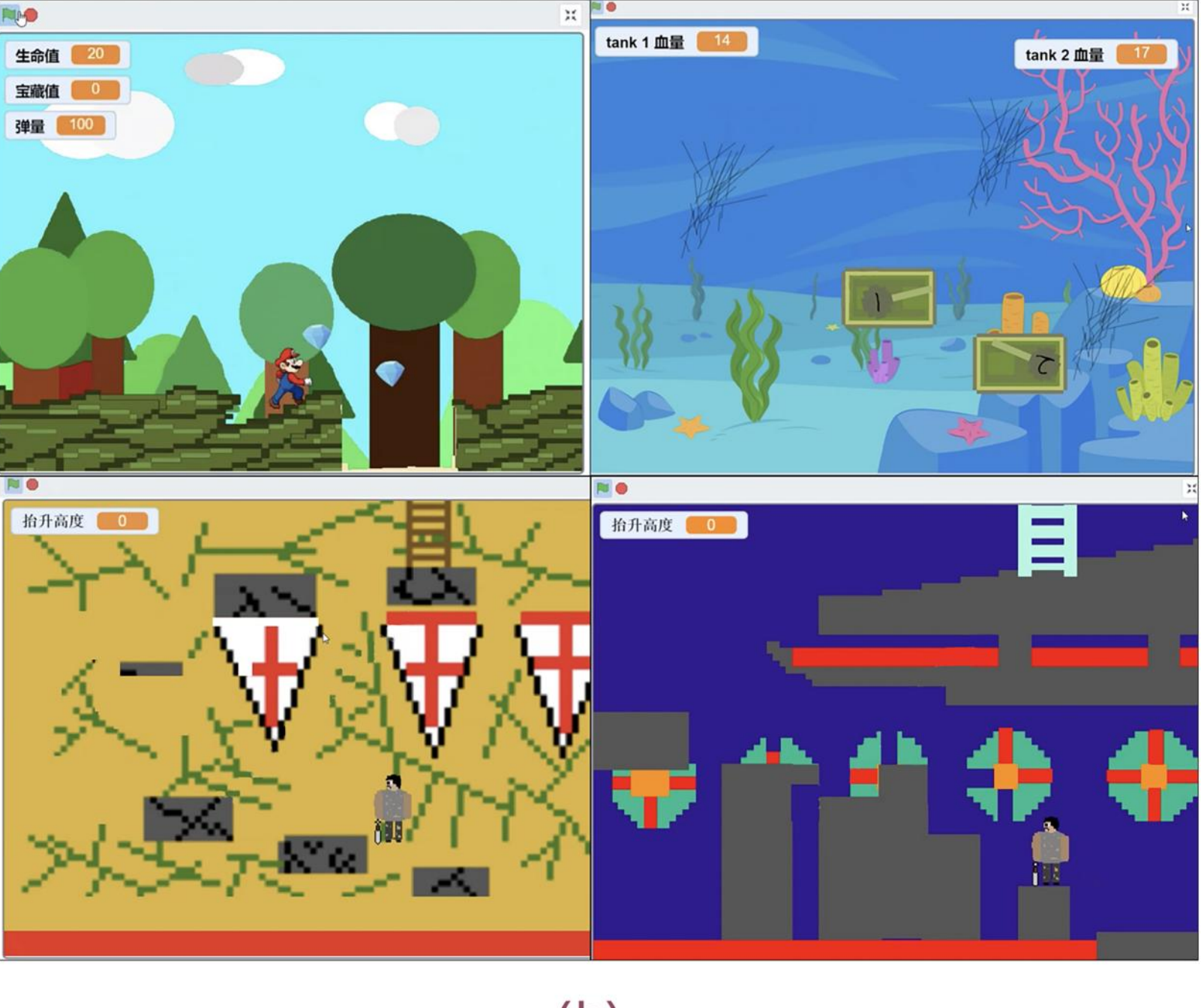

Figure 6. Samples of students' final project. (a) Scratch code of the student-created game; (b) student-made game.

It cannot be denied that there are challenges in the design of course activities. As mentioned by P1: *"There is a huge disparity in team members' abilities, which makes it difficult to speak up. Some have good input, while others are too disorganized."* Younger children may face learning difficulties, as expressed by P14: *"There are significant differences in children's foundational knowledge. Younger children may struggle to keep up, affecting their confidence."* These drawbacks may hinder students' motivation to learn, as described by P17: *"Many children hesitate to ask questions because they feel that other children are too proficient."*

The course design in this project incorporates scaffolding, making it explicit and tightly integrating learning objectives with task design to provide students with a clear understanding of the purpose and significance of each task. This scaffolding helps students grasp the core concepts and skills of the course by providing them with structured support and guidance, enabling them to progress in their learning. Furthermore, the clarity provided by this scaffolding gives students a clear learning direction, enhancing their overall learning experience. As mentioned by P1,

> "This course is designed comprehensively, allowing students to understand and comprehend the structure rather than simply completing tasks. It is different from other programming courses as it explains why things are done in a certain way and how games are designed."

Although the task design has received many positive comments, we still found that students had a weak sense of team identity during the learning process. For instance, students tend to prioritize their individual rankings over the team's rankings, suggesting that this mechanism may not significantly enhance collaborative efforts. As stated by P21,

> "I'm not sure how significant the team aspect is. Students seem to be less concerned about the team ranking and more focused on their personal ranking."

This can also be observed in the task completion data we have collected, as shown in Figure 4. When team tasks and individual tasks are assigned simultaneously, the completion rate of team tasks is consistently lower compared to individual tasks. During interviews, students and parents occasionally mention that group collaboration is not smooth. As mentioned by P1, *"Online, students are unfamiliar with each other, and there is a lack of team cohesion."* The lack of team cohesion poses a challenge to the execution of team tasks.

In this project, we also introduced gamification methods by developing a leaderboard. This reward mechanism to some extent motivated students' learning. As mentioned by P21,

> "Children tend to rely on external rewards to some extent. Giving small rewards to highly motivated children is helpful for their learning."

However, some parents have concerns about gamification. They may worry that gamification will not truly motivate children to learn intrinsically, but will instead focus the experience on external incentives, which will have a detrimental effect on children's learning in the long term. As said by P15,

> "I would be more resistant to this gamification mechanism and worried that my child would focus on outside stimuli. I would want my child to be more intrinsically motivated than externally pursued."

Overall, these design guidelines effectively enhance students' learning experiences. However, there are also certain considerations to be taken into account, such as the design of gamified motivational mechanisms, which require further in-depth research.

### 5.2. Communication channels: WeChat-centered multiple platforms

In our program, students were actively engaged across a variety of platforms to meet the demands of online communication and collaboration, as illustrated in Figure 5. As depicted in Figure 7, our online collaborative learning program integrates four principal types of apps and platforms (see Supplementary Appendix C): the Xiaoe-tech learning platform; the WeChat instant messaging app; the Voov Meeting video conferencing app; and STEAM creative support tools, which include the Scratch programming tool, the Piskel digital painting tool, and the GarageBand music production app. Xiaoe-tech serves as the primary platform for live streaming, releasing recorded courses, and distributing various tasks. Teachers upload pre-recorded videos and associated assignments to Xiaoe-tech, enabling students to view the recorded course and practice the learned material simultaneously.

Of all the communication channels in use, WeChat emerged as the central hub in our collaborative learning program. It facilitated a myriad of interactions: teachers utilized it to address inquiries from children and parents; teaching assistants employed it to remind students about their lessons and tasks; and parents, as well as students, used it to pose questions, discuss topics, share study resources, and coordinate online meetings. The group chat function of WeChat was predominantly used for connecting and communicating. We established both large official groups (encompassing all parents, children enrolled in the course, teachers, and course assistants) and 17 smaller official groups (created for groups of three or four same-aged students to facilitate collaborative communication, with their parents also included in these chats). Additionally, some students took the initiative to create their own unofficial small groups.

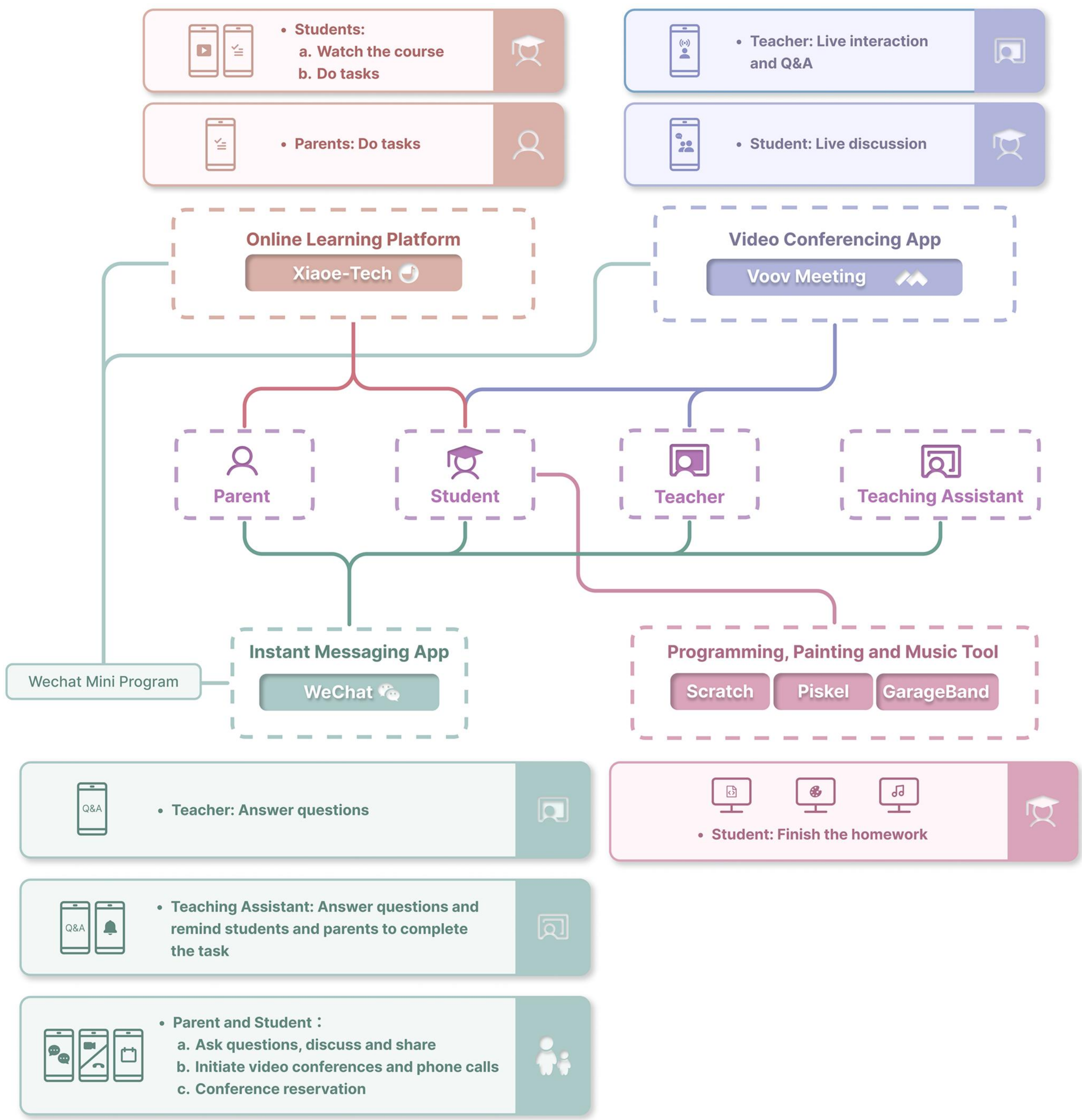

**Figure 7.** Teachers, parents, students, and teaching assistants interact with multiple online learning platforms.

The most important feature of WeChat is that students' actions on it are recorded in the form of "chat history," which can be accessed at any time, making it easier to recall collaborative memory. This benefit is further amplified by the fact that our course is related to programming, digital art, and digital music, all of which can be sent and shared in file format. For example, C7 said,

> "Online courses make it easy to share your own and your team members' tasks, and they can store chat logs and presented works. Using computers is more entertaining too, and sending games and photographs is more convenient."

Online learning is a kind of computer-mediated learning, which is particularly suitable for some courses mediated by electronic devices, such as programming and design. Due to online collaborative learning that enabled team members to share their work, students could also watch other students' and teachers' Q & A, and take part in other students' discussions due to the chat record being available in the official group, which encouraged them to participate more actively in online collaborative activities.

Besides, students could share work files for discussion and modification with their team members on the WeChat group as it is a flexible rich media application that enables users to publish information in different formats, including photographs, videos, and text files. For example, in official small group 14, members C1 and C23 actively discuss on completing group team tasks – exchange programming. They tested, modified, and frequently discussed applications within the WeChat group, as shown in Figure 8.

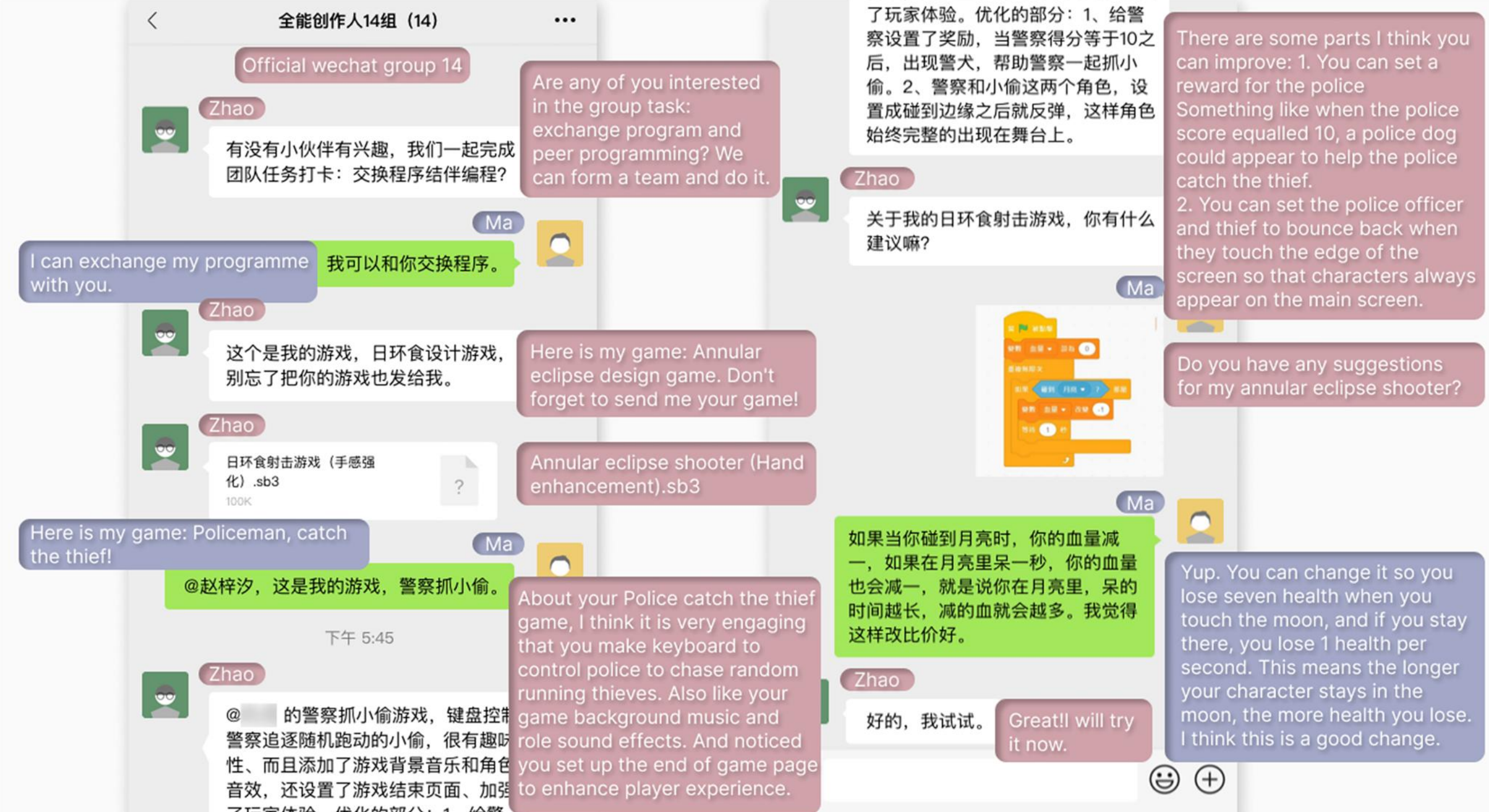

**Figure 8.** WeChat communication and discussion between C1 and C23.

Moreover, WeChat also supports the "WeChat mini program" plug-in. Teachers, students, and parents could access Xiaoe-tech and Voov Meeting functionality directly through this WeChat feature without downloading two apps. WeChat has evolved into a hub connecting other applications and for all the above reasons, was the center of our course organization, communication, and implementation. In sum, we can see the above rich feature makes WeChat as a central channel for communication and collaboration.

While WeChat was essential for communication in our program, its use comes with challenges. The sheer volume of messages can be overwhelming, risking important information being missed. Not everyone may be equally comfortable using the platform, and privacy concerns, especially with children involved, cannot be overlooked. Thus, while WeChat enables real-time interaction, addressing these issues is crucial for effective online collaborative learning.

### 5.3. Benefits: Enabling a new simultaneous multi-thread collaboration

In our program, the first author observed that some children (such as C12 and C23) were especially popular. They would be invited into many other teams' WeChat groups and conduct multi-thread discussions simultaneously. For example, P17 mentioned,

> "Many students and parents will seek the advice of a girl (C12) who was very active in the WeChat group, or even build a new WeChat group for collaboration with her; she could simultaneously interact with others and help them with their questions."

Furthermore, students would actively seek assistance from their excellent peers. For instance, C20 stated,

> "I knew a guy (C23) who was great at programming, and whenever our group was having trouble, I would ask him for help, and a lot of other people asked him too. We had a small group, and sometimes when he was busy answering other groups' questions, I would post my question in the group, wait for his feedback, and then revise my work."

In contrast to offline collaborative learning, online collaboration does not necessitate participation in the entire conversation and follow-up discussion. Delayed responses are acceptable in online collaboration and communication because all participants can read the chat context at any time. Even if they respond later, they can easily pick up where they left off based on chat history. Students can easily send their assignments and questions to a WeChat group, then wait for a response, discuss, and communicate. As a result, when working together online, students can switch between different groups and join and leave discussions more easily, resulting in a multi-threaded collaboration mode. This property can help group members to collaborate more flexibly and freely, thus dividing up the work to accomplish the group's tasks.

Online learning breaks geographical constraints via collaboration. Parents and children can collaborate and communicate online while comparing and promoting one another. In addition, they can fully comprehend students' subject and learning content regardless of their location. In particular, online learning can provide students from areas with limited learning resources the opportunity to collaborate with students from locations with superior learning resources. For example, P17 said,

> "As a parent who lives in a second-tier city, I could follow the development of education in the first-tier city through this program. After seeing the amazing artwork children in Beijing produced, I have a deep belief in my own children's education. My children will benefit from their experience learning alongside children in developed cities for their future development."

One of the benefits of online collaboration is that it allows group members from different locations to communicate and work together without being constrained by the physical space they would otherwise need to share in offline collaboration. Students from throughout the nation engaged well in the official group and freely built new personal groups. Some students built unofficial small groups or communicated one-on-one when they met their team members in official groups. Some students from various groups would form one new group and have a discussion. For example, P17 mentioned,

> "My child's group consisted of four children from various regions of the country. Many children, especially older ones, seem to be passionate about collaborative learning. Therefore, we simply divided the students into a suitable group and allowed them to communicate freely."

Unlike some previous studies, we apply a combination of synchronous and asynchronous online collaboration to a program of children's learning, focusing not only on collaboration between children but also on how parents and teachers are involved in the collaboration process. The collaboration system was also described and behaviour in this study. We discovered that asynchronous communication channels are crucial for supporting children's online collaboration. Children not only communicate asynchronously but also coordinate as a team through scheduled synchronous online meetings. By eliminating time and location constraints, flexible online collaborative learning introduces new collaboration styles, benefiting both children and adults.

### 5.4. Challenges: Hard to initiate and sustain

Although the retention rate of our program is higher than other open online learning reports, there almost two-thirds of children drop out. We found there are several challenges of online collaboration online from our study, which result in online learning's initiating and sustaining problems. The detailed challenges in the four learning phases are listed in Table 2.

When children are first onboarded onto our program, they are required to use a variety of software, which results in a significant technical workload due to the computer-supported collaborative learning involved in online education. The original plan for this program was to use the Xiaoe-tech platform throughout the entire course project. However, the platform's functionality does not fully meet the requirements of children's collaborative learning and is also not sufficiently child-friendly. This was due to the four issues we observed, as listed below:

1. People are not used to checking Xiaoe-tech frequently, and respond to messages more slowly than they would on WeChat.
2. Xiaoe-tech cannot support group discussions like WeChat.
3. Xiaoe-tech's discussion forum for non-concurrent collaboration is rather basic and does not support the upload of source files or in-depth collaboration.
4. Xiaoe-tech's interaction for task submission is too complex for students to complete independently.

For example, P5 said, *"Task submission is too complicated; sometimes it cannot even upload pictures, and we need to toggle back and forth between the software and WeChat."*

With the program going on, the COLP program challenges students' communication and collaboration skills. As shown in Figure 4, it was found that the rate of completion for the first team task was half that of the individual task in week one. We found it difficult to "break the ice" online when the group was newly established and group members did not know each other. P1 said, *"This type of online group collaboration is good and allows children to discuss remotely, but students do not know each other, so there is not enough team cohesion."* It may be that in the absence of social pressure to communicate online, some students and parents would not talk to each other at all no matter what other parents proposed, which may have discouraged active parents. For example, P21 said,

> "At the beginning of the discussion about the team name and logo, I communicated with the other parents, and none of them responded. The parent who did not speak would never speak, and I did not try to communicate with them again."

Moreover, when students are in trouble with the project and need help from their team members, the asynchronous nature of online communication prevents them from getting timely feedback from each other. For example, aligning discussion time between team members was also a significant challenge. As P3 said, *"It's not particularly convenient because you can't meet in person and have to adjust your time."* The above findings show parents and students all find online collaborative learning challenging and inefficient, especially when they encounter trouble and need in-time help.

**Table 2.** Typical challenges at different stages of the COLP program.

| Stages | Typical challenges |
| --- | --- |
| Onboarding | The learning platforms and tools are unfamiliar and not child-friendly for students. |
| Ongoing | Students lack enough communication and collaboration skills. |
| In trouble | Students cannot get sufficient immediate feedback from each other. |
| Overtime | Online learning requires students to have strong self-motivation and self-management skills. |

The retention rate continuously decreased over time because long-term online learning requires students to independently manage their course viewing and task completion time, which presents a challenge to their self-management skills. If students lack strong self-management, planning, and execution abilities, it may result in inefficiency. As mentioned by P14,

> "The biggest difference between online and offline teaching lies in managing the learning pace. Online learning poses more challenges for children because offline classes have a fixed schedule, whereas online assignments and course tasks require children to plan on their own. If they fail to plan well or receive limited parental support, it becomes difficult for them to keep up with the pace."

In summary, students still face numerous challenges in each phase of the COLP program, which is also the reason that online collaborative learning is uniquely hard to initiate and sustain.

### 5.5. Parental involvement and guidance of collaborative skills

Promoting parental involvement is one of our program design guidelines, and we did find it increases parents' workload but also surprisingly improves the children's collaborative skills.

#### 5.5.1. Multiple roles of parents and the increasing workload

Children's online collaborative learning is particularly challenging and requires parental assistance and involvement, parents must play a number of roles in the process, including supervisors and companions. In order to support online collaboration, parents needed to provide social accounts that were required to remain active for extended periods. In our program, we chose WeChat—a universal platform—for communication. However, due to their age, most children did not yet have their own devices or accounts. For example, P23 says, *"I sometimes take on the role of my child's secretary, reminding him about scheduled assignments and his learning progress, as well as assisting him to punch cards and perform other study assistance activities."* P7 added,

> "Playing the child's programming game is what I generally do when we have free time. His dad is especially keen to revise his work when we offer advice based on observations about other children's good work in the group. When a problem cannot be solved, the child will be taught how to communicate with the teacher and how to start group discussions."

In addition, as a result of students' waning motivation due to the non-concurrent nature of online learning, parents frequently needed to encourage and guide their children to actively participate. For example, P21 stated,

> "My son pays lots of attention to the leaderboard. When he lost confidence and patience in the learning process, I always use this ranking to motivate him."

Children's online collaborative learning requires parents to provide technology and tools to support them. Parents frequently needed to share accounts with their children during this project to make it easy for them to participate in collaborative discussions with their peers. At the same time, parents could monitor their children's progress and keep up-to-date with their learning thanks to a new level of parental engagement hardly seen before. Additionally, parents were frequently required to share accounts with their children so they could participate in collaborative discussions with their peers. For example, P20 said,

> "My child has learned a lot during this time, and I have gained a lot from the program too. Parents needed to submit many assignments during the learning process, so I was familiar with my child's homework topics and engaged in the games she produced. Because my child does not have a personal WeChat account, she needs to use mine to communicate with others, and I have been involved throughout the entire process."

In addition, completing tasks can be burdensome for parents due to busyness and failure to keep up with the content of their child's program throughout. P1 mentions, "There is no real-time monitoring and engagement in the child's learning process, and it is difficult to get parents to post comments directly. Participation is challenging and requires parents to learn as well, and we are too busy. Also, the child goes and submits it when he/she is done on his/her own and doesn't communicate with the parents, the assessment is only done on an ad hoc basis every time the group is pushed and the parents are not aware of the situation most of the time."

In summary, we found that parents played multiple roles when assisting their children with online collaborative learning. They also gained a deep understanding of their children's learning process and progress through a new level of involvement never before seen.

#### 5.5.2. Parents' guidance and model of collaborative skills

During children's online collaborative learning, parents frequently provided support and guidance. This had a significant impact on their online collaborative skills. In contrast to offline collaboration, online collaboration frequently occurs at home, where parents can serve as role models and teach their children social and collaborative skills. For instance, when the first team task was uploaded, parents would demonstrate effective collaboration with their children. As depicted in Figure 9, parents took the initiative in instructing children in the group how to schedule appointments, develop meeting topics, and take notes.

In this kind of learning process, parents are able to observe their children's collaborative learning at home, be deeply involved in their children's collaboration, provide timely guidance and behaviour, and influence their children's collaborative learning ability. For example, P2 states,

> "What struck me was that the children would take minutes in the group meeting. Some of the parents suggested regular group meetings and provided the children with a meeting report template from their workplace for reference."

In addition to suggesting a regular meeting system and providing guidance based on their own minute-taking experience, parents also offered suggestions on how to align

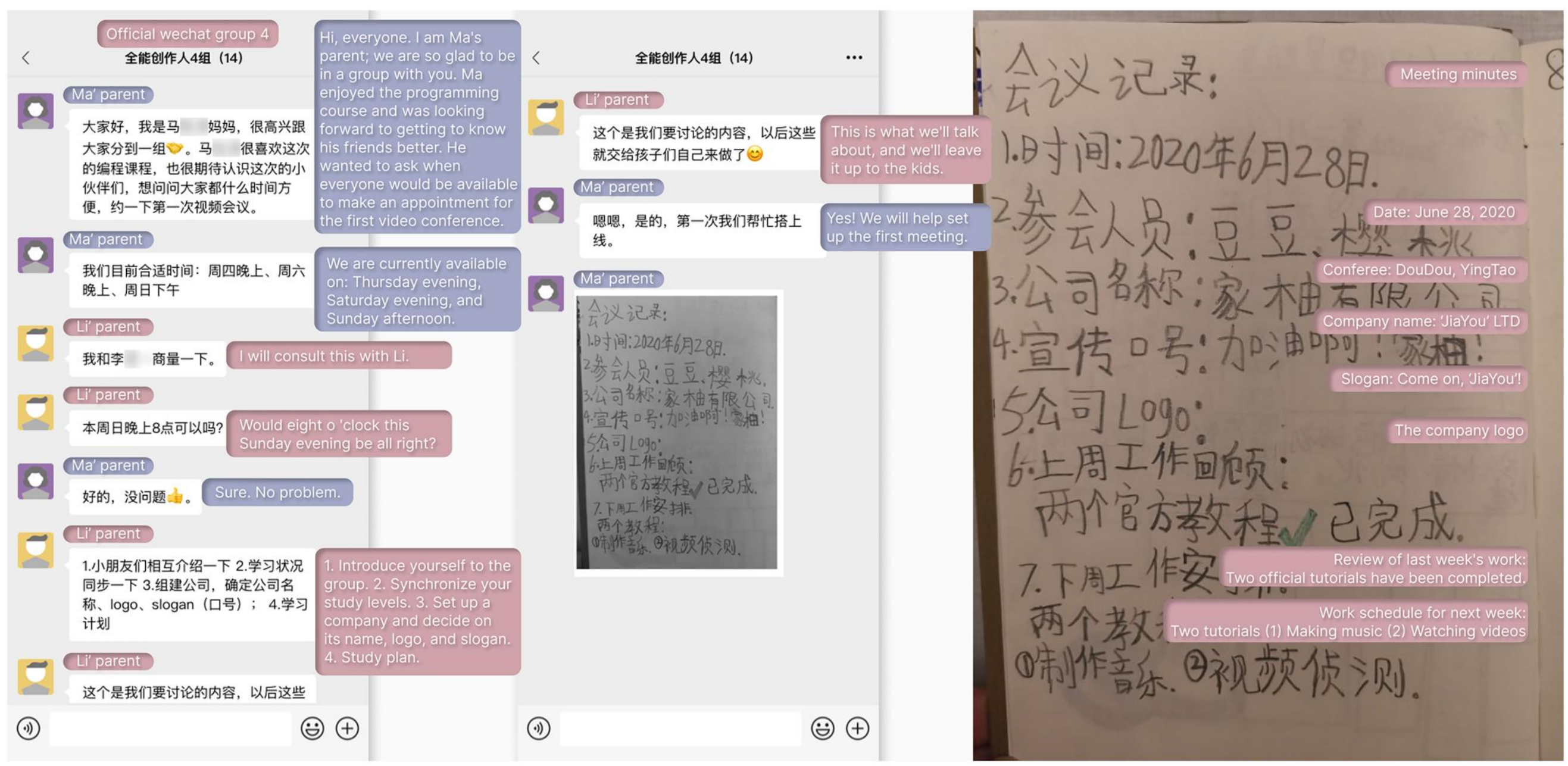

**Figure 9.** Parents' guide their children to have regular meetings and children's group discussion notes.

partner opinions during group discussions. For example, in the interview, P3 said that,

> "At the beginning of the first team task—designing the team logo—everyone had different opinions; then one parent made the wonderful suggestion that everyone should draw their ideas, then merge everyone's icons to make their thoughts together and reach a consensus."

As shown in Figure 10, under this parental guidance and influence, children gradually gained the skill of collaborative learning as they discovered how to incorporate different partner opinions and suggestions. As a result, parents can observe their children's collaboration in detail during the online collaborative learning process, and provide suggestions to develop students' social and collaborative skills. All this parental involvement had a significant impact on children's collaborative learning.

## 6. Discussion

### 6.1. Children's online collaborative learning

Overall, the team task retention rate of our long-term learning project is 36%, which is higher than the previous report (Chuang & Ho, 2016; Simpson, 2013; Zhenghao et al., 2015). It indicated that the program is successful in engaging children with real, practical options for online collaborative learning. Children rapidly accepted and adapted to the online collaborative learning style and benefited from it.

Our online collaborative learning initiatives digitize the collaborative process, while all collaborators can check chat history at any time. This feature has significantly impacted and altered collaborative learning. As long as one user starts a discussion in the group, its content will be available to all group members, allowing them to browse and review the discussion record whenever they like, and take part. Due to the flexibility and asynchronous nature of online collaboration, two collaborators can start, stop, or continue a conversation at any time, or move between various groups or chats, producing a multi-threaded collaboration.

This occurred frequently during our project. Online collaborative learning allows for more flexible communication and interaction between students. They occasionally decided to leave the original group, consult with other people, or start a new one. Some students might be requested to join many groups at once to engage in multi-group, concurrent collaboration, and discussion on various subjects. Ebba Ossiannilsson and Nicolas Ioannides noted that mobile devices, systems, and technologies enable the "anytime" and "anywhere" distribution of information while elaborating on the definition and development of mobile learning (Rennie & Morrison, 2013). However, they skip the specifics of how this impacts teamwork. A remarkable result of our study is the synchronicity brought about by digitizing collaborative processes on multi-threaded cooperation.

One of the main complaints from the community about online learning is its poor participation rate. After organizing earlier research data, Phuong Pham and Jingtao Wang noted that when using significant massive open online course (MOOC) platforms, participation rates were quite low (Pham & Wang, 2018). For instance, edX has a 7.7% retention rate Zhenghao et al. (2015), while Coursera has 4.0% (Chuang & Ho, 2016). However, thanks to the advantages of the aforementioned features, a certain degree of participation was maintained for our project. It is comparatively high in comparison to previous online learning initiatives. In addition, program graduates performed well, demonstrating our online program's strength for collaborative learning.

### 6.2. Parents' role in children's online collaborative learning

Parents play a very important role in children's learning, and parental involvement is a popular topic of discussion.

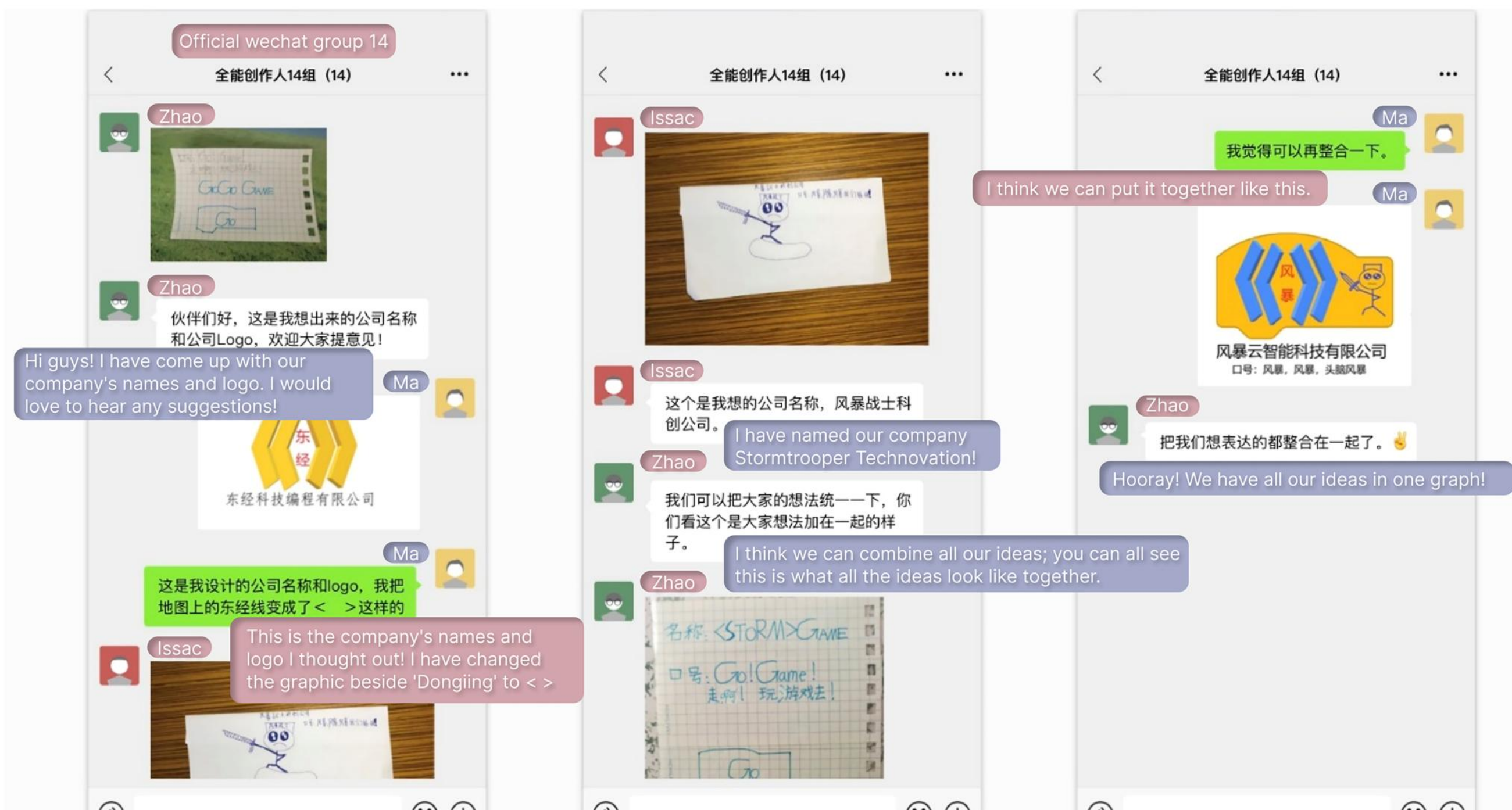

**Figure 10.** Students collaborating on the team logo design.

In previous studies, parents' roles in children's learning have been categorized as audience, implementer, and teacher, and they were involved in children's learning in a variety of ways (Yu et al., 2020).

In our research, we found that parents also play the role of secretary, taking on the responsibility of monitoring, motivating, and guiding their children's online learning. However, in our project, a new role has emerged for parents: that of social developer. Because of the emphasis on online collaboration in our research, a large part of the students' learning was done collaboratively. Collaboration is a complex skill that needs to be developed and routinely practiced in order to be mastered Falkner and Munro (2009).

Although collaborative learning has a positive impact on students, when it does not go well, it can cause unnecessary anxiety and other problems for students, making them less effective learners (Falkner et al., 2013).

In addition, the younger students in our program, who lack collaborative experience and learning skills, had difficulty with online collaborative learning and needed more guidance and collaborative skills. Parents, therefore, needed to be involved in the process, of behaviour and guiding collaborative learning. For example, in our study, when students were engaged in collaborative sessions, some parents would model skills—such as making appointments, developing and conducting discussion, minute-taking, and reaching agreement when there were differences of opinion—for their children. Therefore, in our project, parents were involved in developing collaborative learning skills, behaviour and guiding collaboration, and social interaction.

The following should be brought to the attention of platform designers: when designing online collaborative learning platforms for children, the role of parents should be re-evaluated and innovative content designed with features that help parents guide their children through online collaborative learning. At the same time, additional stakeholders such as teachers, teacher's assistants, and children's extended family members should be consulted to focus on and guide children's collaboration, and provide timely assistance so children can receive careful, targeted guidance and behaviour while learning online.

### 6.3. Design guideline reflections

Based on the results of the project execution, the proposed design guidelines were discussed and refined, and the implications for future course design were summarized.

#### 6.3.1. Prepare learner-centered activities to support online active learning

According to design guideline 1, in the project, we adopted a blended teaching approach that combines live courses, recorded courses, interactive Q & A, and project-based learning, emphasizing the learner-centered design objective. Before the start of the course and tasks, we assess the students' situation to prepare the learning materials accordingly. During the students' learning process, we provide them with materials and scaffold their independent learning. Additionally, we arrange live Q & A sessions and online discussions in WeChat groups to promptly follow up on students' learning progress. During interviews, most parents expressed approval of our project, stating that students actively engage in the course's learning activities. Furthermore, according to statistics, our course has a higher student retention rate compared to other similar online courses.

Therefore, when designing online courses, we recommend emphasizing the learner-centered approach, focusing on students' autonomous learning, and tailoring the course and tasks to their characteristics and needs. Additionally, incorporating various teaching methods such as live courses, recorded courses, and interactive Q & A can provide diverse learning conditions to cater to different learning styles and maximize students' learning motivation. Simultaneously, providing appropriate support and guidance is crucial to ensure that students receive the necessary assistance during their learning.

We also recommend prioritizing a learner-centered approach and considering students' characteristics and needs when setting up an online collaborative learning platform. By incorporating a blend of teaching methods, providing adequate support and guidance, and creating an interactive and user-friendly platform, online courses can offer a rich and engaging learning experience that promotes student success.

#### *6.3.2. Optimize the online group formation method for observational learning*

According to design guideline 2, we designed a children's online collaborative learning environment to support children's observational learning. Through online collaborative learning, students have the opportunity to form teams with peers from diverse backgrounds, knowledge, and skills. By engaging in observational learning, they can effectively learn from others' experiences and perspectives.

Therefore, we suggest that future online collaborative learning should prioritize the group formation method. First, it is important to consider multiple characteristics of students, including their abilities, ages, learning styles, interests, communication, and teamwork skills, and incorporate them into the collaboration process (Moreno et al., 2012). This can be achieved by utilizing more effective group formation methods, such as relevant algorithms and data analysis techniques. Additionally, the unique nature of online courses should be taken into account when forming groups. In summary, it is crucial to introduce more effective online team-building and communication training activities to facilitate communication and collaboration within groups, observational learning within the group should be encouraged.

#### *6.3.3. Involve different roles to form learning communities*

As stated in design guideline 3, parents play an important role in children's learning. Designers can create a collaborative learning environment, building a learning community inside the platform, and providing students with a place to share learning resources and learn collaboratively. This supports the development of open communication between teachers and students, enabling the creation and sharing of information, as well as easier access to other people's learning outcomes and resources. Additional students from around the world may also be attracted to collaborative learning due to the expanded reach of online collaboration groups. The approach to course organization may be similar to a Cmooc. The connectionism-based Cmooc open course approach stresses learning material as the starting point and encourages resource sharing and discussion among students (Joksimović et al., 2015; Siemens, 2004; Wong et al., 2019). The community can also be made aware of projects students have created, allowing other group members and even those not enrolled in the course to vote on and evaluate them.

#### *6.3.4. Personalize tasks and scaffolding to guide diverse students*

According to design guideline 4, we design personalized tasks and scaffolding to guide diverse students. Our research has found that tasks at a single difficulty level are difficult to accommodate due to the diversity of student characteristics. Although the majority of students expressed that our task designs were of moderate difficulty, there were also a small number of students who found the tasks either too easy or too challenging. Therefore, referring to design guideline 4, we suggest personalized design for both group and individual tasks in the future, assigning different tasks based on students' circumstances and providing scaffolding support.

Further, we have found that younger students often have difficulty typing to communicate, and there is a significant gap in the children's ICT skills based on location. Referring to design guideline 4, we, therefore, recommend that children's online collaborative learning platforms develop a child-friendly Jacob and Mythili (2008) input system such as voice input, automatic word correction, predictive text for common phrases, etc. In addition, tangible tools can also be designed to help children collaborate and communicate. This will enable children to communicate without barriers and facilitate their online collaborative learning.

#### *6.3.5. Balance gamification external mechanisms and stimulate students' intrinsic learning motivation*

According to design guideline 5, We introduced the gamification mechanism of the leaderboard. Our gamified learning mechanisms have been well-received by many students and parents. However, we have also found that some parents express concerns about the gamification approach, fearing that external rewards may hinder the development of students' intrinsic motivation and the cultivation of good learning habits.

Therefore, we suggest that gamification design should balance gamification and stimulating students' intrinsic motivation for learning to better guide them to deeper learning Zha et al. (2023). Furthermore, it is important to regularly evaluate and optimize gamified learning mechanisms. By collecting feedback from students, parents, and teachers, we can continuously make improvements, ensuring that the mechanisms consistently stimulate students' learning motivation and provide learning incentives.

#### 6.3.6. Other reflections beyond prepared design guideline

In our project, we found that children encountered some difficulties in using the online learning platform, such as not being able to complete complex interactions independently and not being able to recognize certain platforms. We believe that the online learning platform we used was not easy enough for younger students. Therefore, we suggest that the design of future e-learning platforms should be more child-friendly, taking into account the children among the target users to improve the design of some functions, or even introduce a "child-friendly" mode to simplify the operation of certain complex functions and use images and text in the interface that are more in line with the cognitive level of children. In addition, children encountered some technical problems when using the online learning platform, such as the inability to upload images. We suggest that the learning platform should be further optimized so that young children can independently complete the functions.

Through the project, we verified that our proposed design guidelines are effective to a certain extent. However, it still has some problems, such as teaming, group work, and technical difficulties encountered when using the platform. In the above paper, we have made suggestions for future designs of the design guidelines, and the platform used. We hope that our proposed design guideline reflections will support the design of future online collaborative learning projects as well as online collaborative platforms.

Additionally, we believe that it is feasible to replicate our project by using alternative software tools as substitutes for the ones used in the COLP project. Within our platform, we utilized online communication platforms such as WeChat for synchronous or asynchronous communication. For teaching purposes, we employed learning platforms like Xiaoe-tech. Furthermore, tools like Scratch and Piskel were utilized by children to complete their tasks.

To replicate our research, a combination of other online chat platforms, online learning platforms, and relevant software tools required in the project can be used. This approach would effectively reproduce the instructional outcomes. During the implementation of our project, we encountered some limitations imposed by the software, such as certain functional deficiencies. Therefore, we recommend that in future replication efforts, consideration should be given to selecting software with more comprehensive features.

### 6.4. Limitation and future work

While COLP demonstrates potential in fostering children's long-term online collaborative learning, several limitations remain. Although China's education policies actively promote technology integration, online learning for younger children is not universally accepted worldwide, as some educational systems restrict its use at this age. This may affect the generalizability of our findings, highlighting the need for future research on how varying national education policies influence the adoption and effectiveness of online collaborative learning. Additionally, our participants primarily came from families with strong socioeconomic backgrounds, where parents had sufficient digital literacy and resources to support their children's online learning. However, disparities in access to technology and parental digital competency may impact children's ability to engage effectively in online collaboration. Future studies should examine COLP's accessibility across diverse socioeconomic groups to ensure broader applicability. Moreover, while this study provides valuable qualitative insights, expanding the participant pool, incorporating controlled comparisons, and integrating quantitative assessments will offer a more comprehensive evaluation of COLP's impact. Investigating how different levels of parental involvement influence learning outcomes will also be critical for refining COLP's design and broader implementation.

Future work will focus on scaling COLP to larger and more varied educational settings while refining its adaptability to different learning needs. Implementing structured comparisons, enhancing adaptive task scaffolding, and evaluating AI-supported facilitation could improve student engagement, collaborative dynamics, and long-term participation. In particular, expanding the sample to include participants from different cultural and educational contexts will help mitigate potential cultural bias and strengthen the program's generalizability. Furthermore, exploring strategies to sustain engagement over extended periods, such as personalized feedback mechanisms and improved mentor-student interactions, could strengthen the program's impact. These efforts aim to establish a scalable, inclusive, and evidence-based framework for online collaborative learning among children.

## 7. Conclusion

In this paper, we have explored the potential for children's online collaborative learning via an education intervention, COLP, a 16-week online collaborative learning program, which was designed based on five design guidelines concluded from multiple learning theories. We conducted this program involving 67 primary school students, and more than one-third of the students completed their final teamwork Projects. Despite challenges, COLP benefited students by facilitating simultaneous multi-threaded collaboration. Parents played a critical role in guiding and behaviour their children's collaborative skills. Based on our findings, we deeply discussed design guideline reflections. This study is the first long-term empirical examination of children's project-based online collaborative learning. We hope the far-reaching implications of our findings could inform the design of future online learning initiatives and influence educational policies, thus preparing our children for the future digital workplace through meaningful, project-based learning experiences.

## Acknowledgements

We would like to extend our profound gratitude to the Lab for Lifelong Learning and the Future Laboratory at Tsinghua University for their invaluable support.

## Disclosure statement

No potential conflict of interest was reported by the author(s).

## Funding

This work was supported by the National Natural Science Foundation of China [Grant NO.62441219] and the National Natural Science Foundation Youth [Grant NO.62202267].

## ORCID

Siyu Zha http://orcid.org/0009-0003-0292-5162
Yuanrong Tang http://orcid.org/0009-0008-9714-4321
Jiangtao Gong http://orcid.org/0000-0002-4310-1894
Yingqing Xu http://orcid.org/0009-0003-6304-4262

## About the authors

**Siyu Zha** is a PhD candidate at the Academy of Arts & Design, Tsinghua University. With a background in educational technology and STEM education, her research explores human-AI collaboration, educational human-computer interaction, and the design of AI-augmented creative learning environments.

**Yuanrong Tang** is a graduate student at Tsinghua University and a research assistant at the Tsinghua Institute for Intelligent Industry. Her research focuses on AI agents for wellbeing, conversational agents, and Human-Computer Interaction (HCI).

**Jiangtao Gong** is an Assistant Professor at the Institute for AI Industry Research, Tsinghua University. Her research interests include natural human-AI interaction, cognitive and affective computing.

**Yingqing Xu** is a tenured professor at the Academy of Arts & Design, Tsinghua University. His research interests include the multimodal natural human-computer interaction. He holds a bachelor's degree in mathematics and a PhD in computer graphics.